\documentclass[%
reprint,
nofootinbib, nobibnotes,
amsmath,amssymb, aps,
floatfix, pra,
longbibliography
]{revtex4-2}

\usepackage{nicefrac}
\usepackage{graphicx}
\def\d{\mathrm{d}}
\newcommand{\cO}{\mathcal{O}}
\def\Eqtrapped{(4)}
\def\EqLcubeIntermediate{(17)}
\def\EqL3D{(5)}

\def\AppOnetoone{S.I}
\def\AppCancel{S.II}
\def\AppTransition{S.III}
\def\AppCube{S.IV}
\def\AppCubeSimple{S.V}
\def\AppCylinder{S.VI}
\def\AppRod{S.VII}
\def\App2D{S.VIII}
\def\AppReal{S.IX}
\begin{document}
\title{Mean path length inside non-scattering refractive objects}
\author{Matt Majic} \email{mattmajic@gmail.com}
\author{Walter R. C. Somerville} \email{walter.somerville@vuw.ac.nz}
\author{Eric C. Le Ru} \email{eric.leru@vuw.ac.nz}
\affiliation{The MacDiarmid Institute for Advanced Materials and Nanotechnology, School of Chemical and Physical Sciences, Victoria University of Wellington,
PO Box 600 Wellington, New Zealand}
\date{\today}

\begin{abstract}
It has recently been argued that the geometric-optics mean path length of rays inside a refractive object under Lambertian illumination is independent of the scattering strength of the medium 	[Savo {\it et al.}, Science 358, 765 (2017)]. We here show that it is in fact different in the case of zero-scattering. We uncover and explain the role of trapped ray trajectories in creating this unexpected discontinuity from zero- to low-scattering.
This allows us to derive new analytic results for the zero-scattering mean path length of simple refractive shapes. This work provides a fresh perspective on the study of path length inside refractive objects, with possible applications in for example the study of scattering by large particles or the design of optical systems.
\end{abstract}

\maketitle


Finding the mean chord length for a random distribution of lines in a given object is a natural question in many areas of physics. It is a seemingly complex task from a mathematical perspective, since one should consider the spatial and angular distribution of lines as well as how they intersect the surface of the object. For convex bodies the answer is however surprisingly simple; given by the mean chord length theorem, which has been known for more than a century~\cite{1884Czuber}. It states that the mean chord length $\langle C \rangle$ is independent of the shape of the object, and only depends on the ratio of volume $V$ to surface area $\Sigma$ as $\langle C \rangle = 4 V/\Sigma$. Proofs from various perspectives have been given~\cite{1971Kellerer,coleman1969random,de2003average}.
It was only fairly recently shown that this theorem can be generalized further to the study of random walks in diffusive objects.
The mean \textit{path} length theorem \cite{2003BlancoEPL} states that the mean path length is still simply $\langle L \rangle = 4 V/\Sigma$; this is independent of both shape and the scattering/diffusive properties of the medium. The validity extends across many fields as it is valid for any random walk inside an object, and is particularly relevant to geometric optics within a closed scattering medium.
One important condition for this theorem is that the entrance point and initial direction are
uniformly and isotropically distributed, which in optics is equivalent to a Lambertian illumination \cite{1971Kellerer}.

Path length distributions and mean path length are central to the design of many optical systems where a ray optics description can be used.
They can be used for calculating the optical properties of absorbing and scattering media \cite{2015MupparapuOE,tommasi2020invariance}, refractive granular media in pharmaceutical powders \cite{scheibelhofer2018spatially}, for solar cell design \cite{2015Sprafke,sychugov2019analytical,sychugov2020geometry}, random lasing \cite{2008WiersmaNP}, and integrating spheres \cite{1993NelsonAO,2016CalibrationAO}.
Ray tracing can also be combined with diffraction effects to calculate the electromagnetic scattering properties of large particles in models such as the geometric optics approximation and physical optics model \cite{1982RaveyJO,chowdhury1992energy,1993MackeAO,bi2013physical,kokhanovsky1995local,sun2017physical}, or anomalous diffraction theory \cite{ackerman1987absorption,mitchell2000parameterization,xu2003anomalous}. These have for example been applied to the study of ice crystals \cite{1993MackeAO,bi2013physical,sun2017physical} for climate modeling. The zero-scattering mean path length is directly related to the orientation-averaged absorption cross-sections of large absorbing particles \cite{vandehulst1981,bohren2008absorption,kokhanovsky1997integral}.
Path length distributions have also been used to derive the scattering phase function of an object analytically \cite{gille1999small}. 

In most of these applications, the object has a different refractive index than the surrounding medium. Rays are then refracted at the boundary and may also be reflected internally or externally, which may increase the path length of some internal rays. Even then, it has been argued recently \cite{2017SavoSCI,2020TommasiPRA} that the mean path length invariance remains valid for scattering samples, and is simply modified by a factor $s^2$:
\begin{align}
\langle L \rangle=\frac{4V}{\Sigma} s^2\label{Ls}
\end{align}
for any 3D convex body, independent of the scattering properties of the sample.
A justification of Eq.~(\ref{Ls}) was given assuming a thermodynamic equilibrium/equipartition \cite{2017SavoSCI,2020TommasiPRA}, and following energy conservation arguments drawing on the discussion in Ref.~\cite{yablonovitch1982statistical}.
In this letter, we focus on the mean path length in the low-scattering limit. The equipartition
assumption and Eq.~(\ref{Ls}) also apply, but we will show that the zero scattering case is different, resulting in a discontinuous transition between low- and zero-scattering for some geometries. 
This is related to the existence of trapped paths (similar to the whispering gallery trajectories inside a sphere), which cannot be populated in the strict absence of scattering.
To further support this argument, we have derived a number of new analytic expressions for the zero-scattering mean path length inside simple 2D and 3D refractive objects. These demonstrate the discontinuity at zero-scattering and their derivations support the physical interpretations in terms of trapped rays.
Less symmetric geometries are also studied using Monte Carlo ray tracing simulations \cite{2020TommasiPRA}, allowing us to discuss the generality of this discontinuity, its sensitivity to imperfections, and its relevance to applications.

\noindent {\bf Mean path length in a refractive sample.}
\begin{figure}
	\includegraphics[width=10cm,trim={30mm 0 0 0},clip=true]{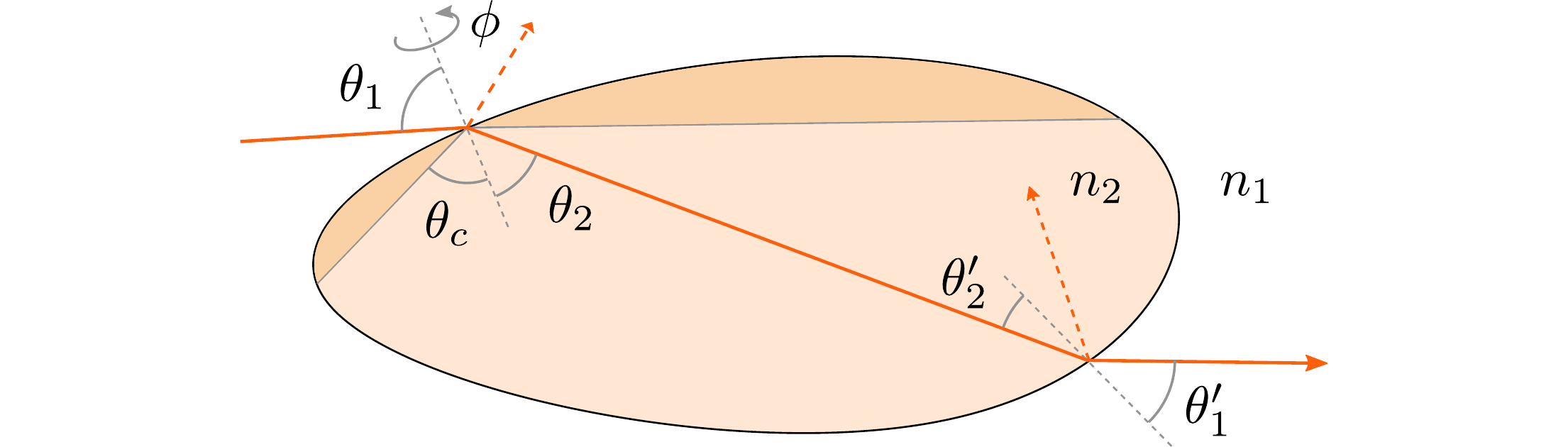}
	\caption{A light ray being refracted as it passes from medium 1 to medium 2. The darker region in medium 2 is not accessible from rays entering from medium 1.} \label{fig snell}
\end{figure}
We consider as shown in Fig.~\ref{fig snell} a non-absorbing convex body $V_2$ embedded in an outside medium $V_1$ with refractive indices $n_2>n_1$ respectively and define $s=n_2/n_1>1$. At this stage, we do not exclude the possibility that medium 2 is a scattering medium.
We study the trajectory of light rays within the geometric-optics approximation, as they undergo stochastic refraction/reflection at the interface.
Consider a ray incident on the surface, with an angle of incidence $\theta_1$ to the surface normal, and rotated by $\phi$ around it. The ray may be refracted at an angle $\theta_2$ with a probability $T_{12}(\theta_1)$ with $\sin\theta_1 = s\sin\theta_2$ (Snell's law), while the azimuthal angle $\phi$ is unaffected.
By optical reciprocity, the probabilities of transmission at complementary angles are identical: $T_{12}(\theta_1) = T_{21}(\theta_2)$. Similar laws apply to internal rays hitting the object surface with internal $\theta'_2$ and external $\theta'_1$ angles. Since $n_2>n_1$, angles $\theta'_2$ above the critical angle $\theta_c=\mathrm{asin}(1/s)$ have zero transmission: there is total internal reflection (TIR). 
 
As for the mean path (or chord) length theorems, we assume that the illumination of the object from the outside is Lambertian. The surface irradiance is therefore uniform and the incident angles follow the probability distribution $p(\theta_1)=2\cos\theta_1\sin \theta_1=\sin(2\theta_1)$ ($0\le\theta_1<\pi/2$) in 3D \cite{1971Kellerer}. Because Lambertian illumination maximizes entropy and energy is conserved in the scattering process, the ensemble of external rays reflecting and internal rays exiting must also follow a Lambertian distribution (otherwise entropy would have been decreased).
 Then, since incident and outgoing rays have the same distribution, for every internal ray making an angle $\theta'_2<\theta_c$ to the normal 
(and reflected with a probability $p_2=1-T_{21}(\theta'_2)$), there is an incident external ray that is externally reflected at the Snell-matching $\theta'_1$ with the same probability $p_1=1-T_{12}(\theta'_1)=p_2$, 
i.e. there is a one-to-one correspondence between these rays (see Sec.~\AppOnetoone~\cite{SI} for more detail).  This allows us to ignore all internally reflected rays with $\theta'_2<\theta_c$ in mean path length calculations, if we also ignore any reflection from the outside (dashed rays in Fig.~\ref{fig snell}).
These externally reflected rays would normally have zero path length, but if they \textit{were} to transmit inwards, they would have exactly the path length and angular distribution of the rays that reflect from the inside for $\theta'_2<\theta_c$. See also Sec.~\AppCancel~for an explicit example of this cancellation in simple geometries. This is a crucial result for refractive objects, as it simplifies dramatically the calculations. 
Note however that the contribution of rays with $\theta'_2>\theta_c$ (TIRs), whose distribution is not specified, must still be accounted for.
 This result also highlights that Eq.~(\ref{Ls}) for scattering media assumes that these externally reflected rays with $L=0$ are included in the statistics, an important point that was not made explicit in Refs.~\cite{2017SavoSCI,2020TommasiPRA}. The mean path length not counting $L=0$ rays can be simply deduced as $\langle L_{L>0} \rangle = {\langle L \rangle }/{\bar T_{12}}$, where $\bar T_{12}$ is the Lambertian-averaged transmission \cite{duntley1942optical}.
 
Following these considerations, we may express the mean path length in the object as
\begin{align}
\langle L \rangle = \int_\Sigma \frac{\d \Sigma}{\Sigma}\int \frac{\d\phi}{2\pi} \int_0^\frac{\pi}{2} \d\theta_1 L(\mathbf{r},\theta_1,\phi) \sin(2\theta_1). \label{L3D1}
\end{align}
where $L(\mathbf{r},\theta_1,\phi)$ denotes the total ray path length for a given entry point $\mathbf{r}$ and incidence angles, including possible total internal reflections until it reaches the surface with $\theta'_2<\theta_c$ (thanks to the cancellation between internal/external reflections). 
In the absence of scattering and total internal reflections, $L$ then coincides with the chord length $C$.
In the presence of scattering $L$ should be understood as an average over all possible scattering paths, which renders this ray approach difficult.

We have applied this method to several standard geometries in the non-scattering case and obtained analytical results for simple shapes and numerical results for
more complex shapes.
The most surprising outcome is that the zero-scattering mean path length $\langle L^0 \rangle$ is different (smaller) than 
the scattering mean path length $\langle L \rangle$ for some geometries, which results in a discontinuous transition at zero-scattering.
We first discuss further this counter-intuitive result as it will provide physical insight into its origin
and provide an alternative method of calculating $\langle L^0 \rangle$ in special cases.


\noindent{\bf Transition between the low- and zero-scattering regimes.} 
To understand how this discontinuous transition arises, we will attempt to connect the two different approaches 
for the scattering (thermodynamic/equipartition) and non-scattering (ray optics) cases. Specifically, we here derive the mean path length $\langle L\rangle$ in the low-scattering limit from $\langle L^0\rangle$ using a ray-optics argument.
At the center of this discussion  is the existence of trapped rays. These undergo successive TIRs and cannot escape, similar to propagating modes in an optical fiber or whispering gallery modes in dielectric spheres. Note that these trajectories may be repeating (as the optical modes) or chaotic. Because of reciprocity, these rays cannot be excited from outside in the ray-optics framework and therefore are irrelevant to $\langle L^0\rangle$.
\footnote{Note that arbitrarily long path lengths may still exist,
for example in the 2D ellipse, where rays that refract in at almost
the critical angle at the tips will undergo a large number of total
internal reflections before inevitably refracting out. In terms of
the elliptical billiard table \cite{1981BerryEJP}, the rays would enter at a trough on
the phase portrait that touches the line at the critical angle.
However, these are not strictly trapped.}
 But if scattering is present, then there is a probability that some rays are scattered into and out of trapped trajectories. For very low scattering, this probability is small and one might expect that it does not affect the mean path length. However, trapped rays exhibit very long path lengths because scattering is low. It is this product of a small probability by a large path length that may result in a finite, non-zero contribution to $\langle L\rangle$ even in the limit of zero-scattering, but not for zero-scattering, hence the discontinuity.

To be more quantitative, we denote the scattering coefficient $\alpha$ and assume that the scattering mean free path $l_\alpha=1/\alpha$ is much greater than $\langle L^0\rangle$, i.e. low-scattering limit. For simplicity, we will here consider special cases where the probability of scattering into a trapped trajectory, denoted $P_T$, is independent of the position of the scattering event. A more general case is discussed in Sec.~\AppTransition.
The average probability of a ray scattering is $\alpha \langle L^0\rangle \ll 1$. Since the average path length for non-trapped rays is of order $\langle L^0\rangle$, scattering into them results in negligible changes to path length (of order $\alpha \langle L^0\rangle^2$). In contrast, a ray will only escape a trapped path if it is scattered again, which results in an average path length $l_\alpha \gg \langle L^0\rangle$ for trapped rays. Moreover, scattering may occur into another trapped path with probability $P_T$, which increases the path length by $l_\alpha$ again until the next scattering event. Summing, we obtain the mean path length for trapped rays as:
\begin{align}
\langle L^T\rangle = l_\alpha + P_Tl_\alpha+P_T^2l_\alpha+\ldots = \frac{l_\alpha}{1-P_T}.
\end{align}
$\langle L^T\rangle\gg\langle L^0\rangle$ but this is compensated by the small probability $\alpha \langle L^0\rangle P_T$ of scattering into a trapped trajectory.
We can now add this contribution to $\langle L^0 \rangle$ to obtain the low-scattering mean path-length:
\begin{align}
\langle L\rangle \approx \langle L^0\rangle + \left[\alpha \langle L^0\rangle P_T \right]\langle L^T\rangle = \frac{\langle L^0\rangle}{1-P_T}. \label{trapped}
\end{align}
This derivation provides an explanation for the discontinuity at zero-scattering, which is due to the second term, related to trapped trajectories.
Eq.~(\ref{trapped}) moreover provides a simple method of deducing $\langle L^0\rangle$ analytically for objects where $P_T$ is independent of position and angle, which includes many objects with faceted sides. An important special case is for objects where no trapped rays can be supported ($P_T=0$), for which $\langle L^0\rangle=\langle L\rangle$. Amongst these are objects with a smaller refractive index than the embedding medium ($s<1$).
The consideration of trapped paths also suggests a link between this problem and the theory of ``billiards'' in classical mechanics \cite{1981BerryEJP}. In particular, ergodic shapes will also automatically have $P_T=0$ (since every ray samples the entire phase space) and therefore $\langle L^0\rangle=\langle L\rangle$.

\begin{figure}
	\includegraphics[width=8.4cm, trim={35mm 0 0mm 0},clip=true]{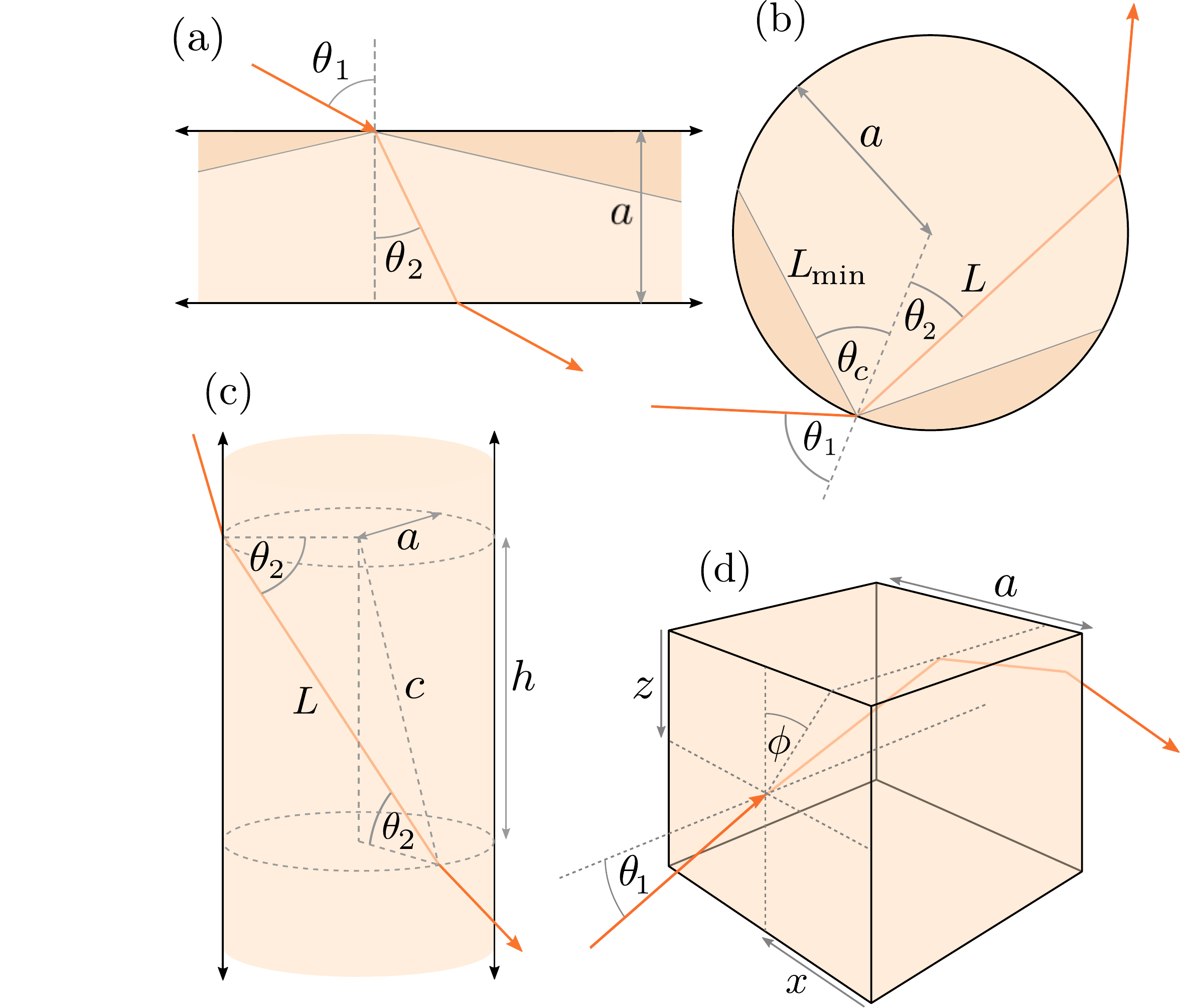}
	\caption{Main geometries considered in this work, (a) the 2D strip and 3D slab, (b) the circle and sphere, (c) the infinite cylinder and (d) the cube and cuboid. We also treat the square, rectangle and infinite square rod.} \label{Fig3Dschem}
\end{figure}

\begin{figure}
	\includegraphics[width=8.7cm]{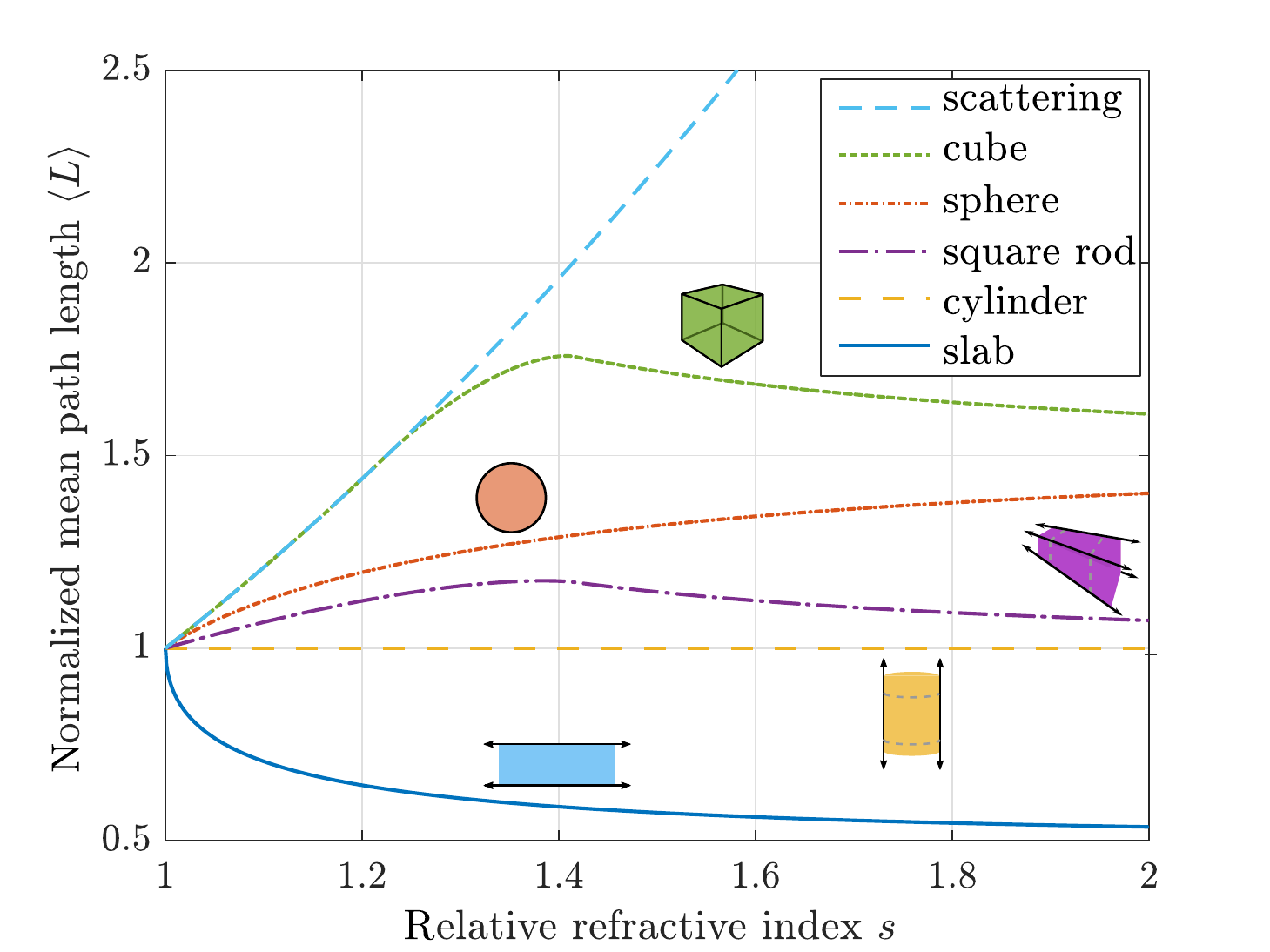}
	\caption{Comparison of the $s$ dependence of the zero-scattering mean path length $\langle L^0 \rangle$ for 3D objects for which analytical expressions were derived. All values are normalized to the mean chord length $\langle C \rangle=4V/\Sigma$.
	The scattering case $\langle L\rangle = s^2 \langle C \rangle$ is shown as a dashed line.} \label{all3Dobjects}
\end{figure}

\noindent {\bf Analytic results.}
To further illustrate this discussion, we now provide a collection of newly derived analytic results for $\langle L^0\rangle$ in simple 3D and 2D geometries.
The main 3D geometries that we considered are summarized in Fig.~\ref{Fig3Dschem}, where their parameters are defined.
The advantage of these ideal geometries is that the derivations illustrate how concepts such as trapped rays affect the mean path length.
The derived $\langle L^0\rangle$ as a function of $s$ for all 3D geometries are summarized and compared to the scattering case in Fig.~\ref{all3Dobjects}.
To calculate $\langle L^0\rangle$, we use Eq.~(\ref{L3D1}), rewritten in terms of the inside angle as:
\begin{align}
\langle L^0 \rangle = s^2\int_\Sigma \frac{\d \Sigma}{\Sigma} \int \frac{\d\phi}{2\pi}  \int_0^{\theta_c} \d\theta_2 L(\mathbf{r},\theta_2,\phi) \sin(2\theta_2) . \label{L3D}
\end{align}

We start with the simplest case of an infinite slab of width $a$, Fig.~\ref{Fig3Dschem}(a). In this case, $L(\mathbf{r},\theta_2,\phi)$ only depends on $\theta_2$.
Moreover, since we can ignore probabilistic reflections 
and it is not possible to excite TIRs from the outside, $L$ is the same as the chord length so we have 
$L = a/\cos\theta_2$
and the mean is calculated as 
\begin{align}
\langle L_\text{slab}^0\rangle 
=2as^2(1- \cos\theta_c) = 2as^2\left(1-\sqrt{1-\frac{1}{s^2}}\right). \label{Lslab}
\end{align}
This could also have been deduced from Eq.~(\ref{trapped}) since the trapping probability is uniform: $P_T=\cos\theta_c$.
$\langle L_\text{slab}^0 \rangle$ decreases with $s$ (see Fig.~\ref{all3Dobjects}) and is less than the mean path length for a slab with scattering,
$\langle L_\text{slab}\rangle = 2as^2$.

For a sphere of radius $a$, Fig.~\ref{Fig3Dschem}(b), $L(\mathbf{r},\theta_2,\phi)$ again depends on $\theta_2$ only and it is not possible to excite TIRs from the outside, so we have $L=2a\cos\theta_2$ (the chord length) and integrating Eq.~(\ref{L3D}):
\begin{align}
\langle L_\text{sphere}^0\rangle 
=&\frac{4a}{3}s^2\left[1-\left(1-\frac{1}{s^2}\right)^{\nicefrac{3}{2}}\right]. 
\end{align}
This expression appears for example in the absorption cross section for large weakly absorbing spheres \cite{bohren2008absorption,kokhanovsky1995local}. For comparison, the mean path length for a sphere with scattering is $\langle L_\text{sphere}\rangle = (4/3)as^2$.
We cannot here use Eq.~(\ref{trapped}) because the probability of trapping $P_T$ depends on position: trapping is more likely for scattering events close to the sphere surface.

For the cube, Fig.~\ref{Fig3Dschem}(d), there are three regimes depending on $s$. 
First, we can show that for $s\leq\sqrt{3/2}$, no trapped rays exist, hence $P_T=0$ and
\begin{align}
	\langle L_\text{cube}^0(s\leq\sqrt{3/2})\rangle = \langle L_\text{cube}\rangle =\frac{2}{3}a s^2.
\end{align}
For $s\geq\sqrt{2}$, we may use Eq.~(\ref{L3D}); the problem is simplified by the fact that all rays exit the opposite face of the cube. Some will total-internally reflect on an adjacent face (see Fig.~\ref{Fig3Dschem}(d)), some will exit straight through, but in both cases, the path length is given as $L=a/\cos \theta_2$, therefore we obtain
\begin{align}
\langle L^0_\text{cube}(s\geq\sqrt{2})\rangle=2as^2\bigg (1-\sqrt{1-\frac{1}{s^2}}\bigg).
\end{align}
In the intermediate case, $\sqrt{3/2}<s<\sqrt{2}$, calculating $\langle L_\text{cube}^0 \rangle$ via Eq.~(\ref{L3D}) is rather technical, see Sec.~\AppCube. 
We present in Sec.~\AppCubeSimple~ a simpler derivation using Eq. \eqref{trapped}, which applies because the trapping probability $P_T$ is again independent of the location of the scattering event. Both result in:
\begin{align}
&\langle L_\text{cube}^0(\sqrt{3/2}<s<\sqrt{2}) \rangle =\\
&{\frac{4as^2}{\pi}}\left( \sin^{-1}(s^2-1) -  \sqrt{1-\frac{1}{s^2}}~\sin^{-1}(2s^2-3)\right). \label{LcubeIntermediate} \nonumber
\end{align}
A similar derivation can be carried out for a cuboid of edges $a,b,c$ and the results are the same with the transformation:
\begin{align}
a\rightarrow \frac{3abc}{ab+bc+ca},
\end{align}
i.e. rescaled by the relative factor $V/\Sigma$ for each shape.
The cases of an infinite circular cylinder (Fig.~\ref{Fig3Dschem}(c)) and an infinite square rod are also derived and discussed in Secs.~\AppCylinder~and~\AppRod.

Finally, 2D objects can be treated with a similar approach. We have obtained analytic expressions for $\langle L^0 \rangle$ for an infinite strip, a circle, a square, and a rectangle. The results and derivations are provided in Sec.~\App2D, along with a graphical summary. The conclusions are similar to those for 3D objects.

\begin{figure}
	\includegraphics[width=9.2cm]{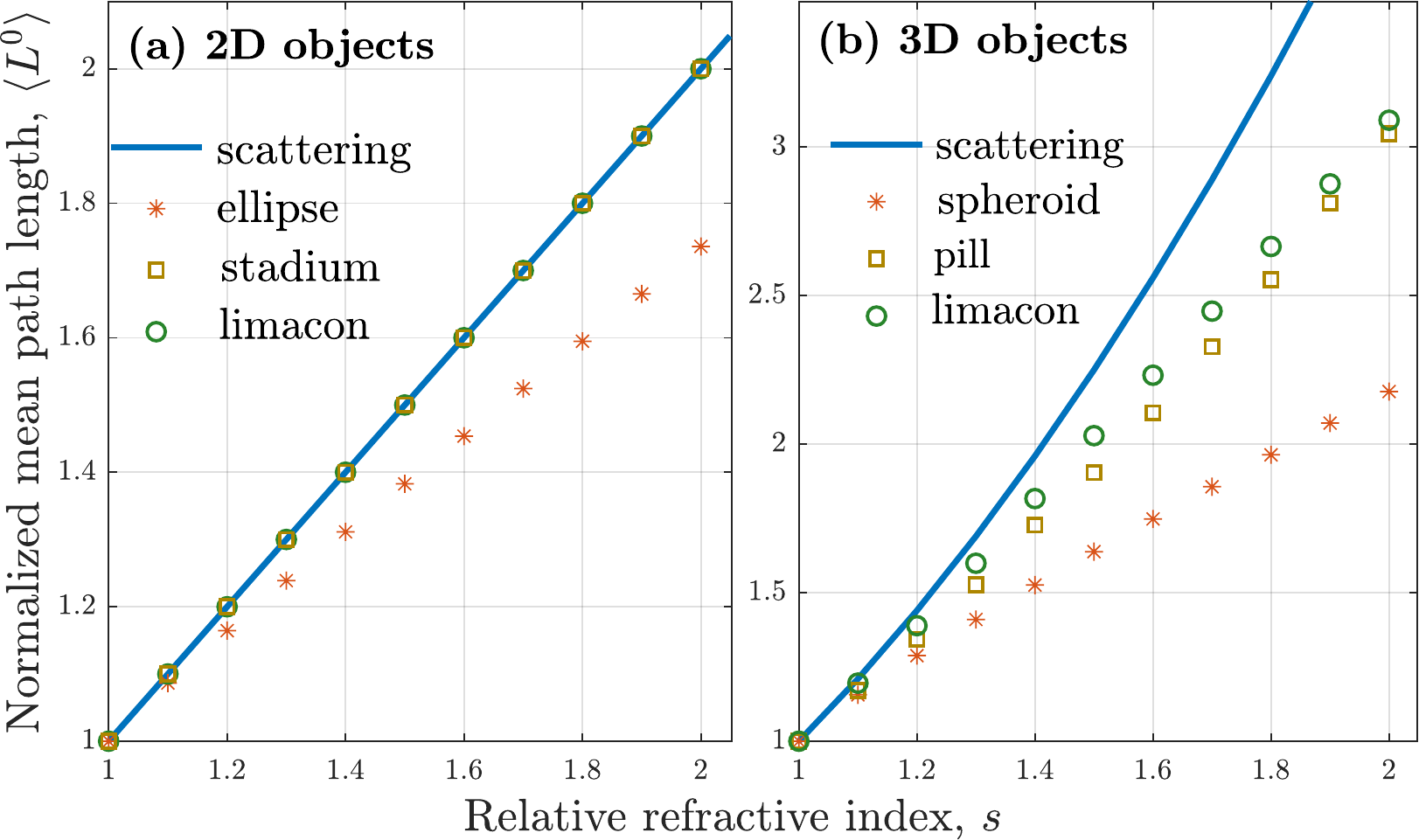}
	\caption{Comparison of the $s$ dependence of the zero-scattering mean path length $\langle L^0 \rangle$ for less symmetric objects.
	All values are normalized to the mean chord length $\langle C \rangle$, and the scattering case ($\langle L\rangle$) is shown as a solid line.
	(a) 2D objects: ellipse and stadium of aspect ratio 2, Pascal's Lima{\c c}on of polar equation $r(\theta) = b+a\cos\theta$, with $b/a=3$. (b) 3D objects: same as 2D objects with symmetry of revolution around $z$.} \label{FigComplex}
\end{figure}

\noindent{\bf Application to physical systems.}
To investigate the generality of these results, we now consider less symmetric geometries, for which numerical calculations (Monte Carlo ray tracing \cite{2020TommasiPRA}) can be used to derive $\langle L^0 \rangle$. Figure~\ref{FigComplex} summarizes these results. We consider explicitly in Fig.~\ref{FigComplex}(a) 2D shapes with different refractive index and decreasing symmetry: ellipse, stadium, and a convex Lima{\c c}on. The latter two do not show any discontinuity within our numerical accuracy, i.e. $\langle L^0 \rangle=\langle L \rangle$, even at high refractive index, which suggests that the probability of scattering into a trapped trajectory is zero. Note that trapped rays may still exist (such as the one depicted for the stadium in Fig.~\ref{FigComplex}(a)), but they correspond to unstable orbits with vanishingly small probabilities of scattering into. The situation is different for ellipses where a larger number of rays may be trapped, in agreement with the theory of elliptical billiards \cite{1981BerryEJP}.
Interestingly, corresponding 3D objects with symmetry of revolution all show $\langle L^0 \rangle<\langle L \rangle$, likely because of the trapped trajectories in the planes perpendicular to the revolution axis.
It would be interesting to further link these results to the theory of classical billiard and chaos/ergodicity but this is outside the scope of this letter. 

Figure~\ref{FigComplex} overall suggests that the mean path length discontinuity is a special property of geometries with higher symmetry.
These special shapes are nevertheless commonly used as model systems in many applications. As an example, ice crystals are often taken as high-symmetry objects to derive their optical properties for atmospheric models \cite{1993MackeAO,bi2013physical,sun2017physical}. Within the geometric-optics approximation \cite{1993MackeAO,kokhanovsky1995local}, the absorption cross-section $C_\mathrm{abs}$ of weakly absorbing objects is directly proportional to the {\it zero-scattering} mean-path-length. The expressions we obtained (and more that could be derived using the same approach) can then be used to derive an analytic expression. For example, for a 5\,$\mu$m-wide ice cube at $\lambda=1\,\mu$m ($s=s'+is''=1.3+1.6\times 10^{-6}i$), we find that the analytic prediction:
\begin{align}
&\langle C_\mathrm{abs} \rangle \approx \frac{4\pi s''}{\lambda}\langle L_\text{cube}^0(s') \rangle \nonumber\\
 & \approx \frac{16as'^2s''}{\lambda}\left( \sin^{-1}(s'^2-1) -  \sqrt{1-\frac{1}{s'^2}}~\sin^{-1}(2s'^2-3)\right)
\end{align}
agrees within $\pm 10\%$ with numerical calculations. This approach is valid for particle sizes much larger than the wavelength, but smaller than the characteristic absorption length, i.e. $1\ll (2\pi/\lambda)a \lessapprox (1/s'')$.
Together with the approximate extinction cross-section $\langle C_\mathrm{ext} \rangle\approx 3a^2 $, these provide simple analytical inputs for atmospheric models over a large size range, replacing the time-consuming ray-tracing simulations otherwise required. This approach could be generalized to more realistic ice-crystal shapes and to other weakly absorbing atmospheric aerosols.

Apart from such applications, whilst the zero-scattering discontinuity is interesting from a fundamental point of view, we should also consider its relevance to real physical systems. Firstly, all physical media are imperfect and should exhibit a small, but non-zero, scattering coefficient. In addition, surface imperfections are unavoidable, be it roughness or small deviation from ideal shapes (likely to make the object non-convex). Thirdly, the ray-optics description is only an approximation and wave effects can affect reflection/refraction, in particular resulting in a small probability of outcoupling during TIR events, which would preclude the existence of strict trapped trajectories. Because of these effects, one could argue that the zero-scattering discontinuity is irrelevant and the general formula (Eq. (\ref{Ls})) applies instead. However, one should also consider that any physical medium has non-zero absorption coefficient. This small absorption negates the contribution of extremely long-lived trapped rays for low scattering, so that the relevant experimental mean path length is in fact the zero-scattering one, as long as the scattering is smaller than the absorption coefficient. This argument is developed more qualitatively in Sec.~\AppReal.

\noindent{\bf Conclusion}.
We have examined how shape affects the mean path length of rays in non-scattering refractive objects, providing the theoretical groundwork to derive the mean path length analytically, and applying it to simple shapes. Crucial to being able to derive these results was the fact that all probabilistic reflections below the critical angle can be discounted if they are ignored from both the inside and outside.
We believe that some other geometries will be able to be treated using the same approach.
These analytic results also highlight explicitly the discontinuous transition from non-scattering to scattering media and
demonstrate that it is due to the existence of trapped trajectories that cannot be occupied without scattering.

We believe this work is an important contribution to the resurgent study of path length invariance in media.
The derived analytic expressions will also be useful in other theoretical contexts where mean path length, or path length distributions, are studied, as refractive non-scattering objects are central to many applications.

\begin{acknowledgments}
The authors are grateful to the MacDiarmid Institute, NZ, for financial support.
\end{acknowledgments}

\renewcommand{\thepage}{S\arabic{page}}

\clearpage
\twocolumngrid

{\center \textbf{SUPPLEMENTARY INFORMATION}}

\section{Derivation of reflection cancellation}\label{AppOnetoone}
\begin{figure}
	\includegraphics[width=8.5cm]{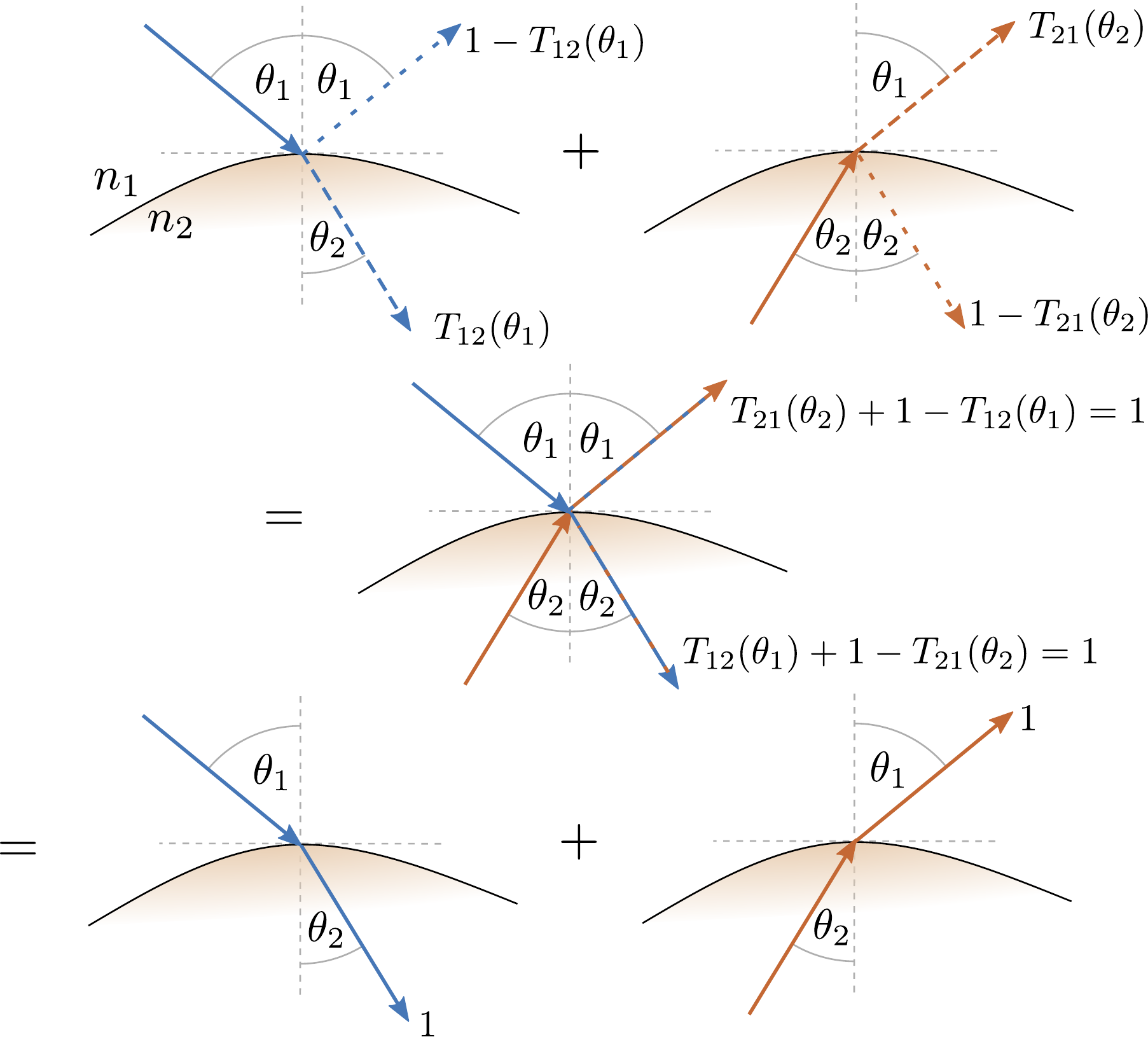}
	\caption{Schematics illustrating the effective cancellation of external and internal probabilistic (non-TIR) reflections. 
		Top: the possible paths of external (blue) and internal (red) rays incident on the surface, where $\theta_1$ and $\theta_2$ are Snell-matching angles. The dashed lines represent the fractional probability of the outcomes. Center: the addition of these two cases, where the origins of the rays are indicated by the coloring of the dashes. 
		Bottom: equivalent scenario where probabilistic reflections are ignored, which leads to the same total number of rays leaving at each angle, so preserves the mean path length.
	}
	\label{FigSchemCancel}
\end{figure}

In the main text we used the fact that all probabilistic (non-TIR) reflections can be ignored when calculating the mean path length. Here we provide a more detailed explanation of this phenomenon with diagrams (Fig.~\ref{FigSchemCancel}).
First we noted that due to retaining entropy, there is a one-to-one correspondence between ingoing and outgoing rays: for each ingoing external ray, there is an outgoing ray at the same angle at that surface point. This in turn means that for each ray incoming at an angle $\theta_1$ to the normal, there is an internal ray headed toward the surface at the Snell matching angle $\theta_2$ where $\sin\theta_1=s\sin\theta_2$. In terms of Fig. \ref{FigSchemCancel}, this means that the red and blue solid lines represent the same density or number of rays.
As shown in the top diagrams in Fig.~\ref{FigSchemCancel}, for each externally reflected ray at an angle $\theta_1$, which was reflected with a probability $p_1=1-T_{12}(\theta_1)$, there is an internally reflected ray, which was reflected inward at $\theta_2$ with the same probability $p_2=1-T_{21}(\theta_2)=p_1$. This is because optical reciprocity ensures $T_{12}(\theta_1)=T_{21}(\theta_2)$.
These cases are added together in the center diagram of Fig. \ref{FigSchemCancel}, showing that the outgoing rays (red with blue dots) are composed of external rays undergoing reflection (with probability $p_1=1-T_{12}(\theta_1)$), or internal rays undergoing refraction (with probability $T_{21}(\theta_2)$), while the ingoing rays (blue with red dots) are composed similarly. In total there is a relative density of 1 outgoing at $\theta_1$ and a density of 1 ingoing at $\theta_2$. 
As shown in the lower diagrams of Fig.~\ref{FigSchemCancel}, the situation is equivalent to a scenario where both the internal and external reflections are replaced with refractions, because this results in the same density of 1 for outgoing and ingoing rays. We say equivalent meaning that the number density of rays is conserved, as well as ray angles and therefore also the mean path length.
So in analytic derivations only reflections for internal rays with $\theta'_2>\theta_c$ (TIR) need to be considered, as these reflections have no correspondence with external reflections.

\section{Explicit example of reflection cancellation}\label{AppCancel}

We can demonstrate the cancellation between internal and external reflections explicitly for simple geometries where the chord length and inside angle $\theta_2$ of a light ray after each internal reflection is identical -- these include the 2D infinite strip, circle, 3D infinite slab, infinite cylinder, and sphere.
In these shapes, a light ray may internally reflect $n$ times, each time with a probability $1-T_{21}(\theta_2)$ so the path length is increased by a factor
\begin{align}
	f=\sum_{n=0}^{\infty}\left[1-T_{21}(\theta_2)\right]^n= \frac{1}{T_{21}(\theta_2)} \label{f sum}
\end{align}
The fraction of rays incident with an angle $\theta_1$ transmitted into the sample is $T_{12}(\theta_1)=T_{21}(\theta_2)$, thereby canceling the factor in Eq. \eqref{f sum}.

\section{Effect of non-uniform probability of trapping $P_T$}\label{AppTransition}

In an object like a sphere, the density of trapped trajectories varies and is larger closer to the surface. The trapping probability $P_T$ then depends on the location of the scattering event. We may still define an average trapping probability $\bar{P}_{T,1}$. But for any ray that gets ``trapped'', there is also a probability that the next scattering event will result in another trapped ray, and so on. If $P_T$ is non-uniform, these subsequent average probabilities $\bar{P}_{T,n}$ may be different to that of the first trapping event. For a sphere for example, one expects $\bar{P}_{T,2}>\bar{P}_{T,1}$ as the scattering event for a ray already in a trapped trajectory is more likely to occur closer to the surface.
Let $P_n$ be the average probability that a ray gets trapped exactly $n$ times, with corresponding average path length $nl_\alpha$.  Then the mean path length may be approximated in the low scattering limit by 
\begin{align}
	\langle L \rangle = \sum_{n\geq0}P_n \langle L_n \rangle. \label{llowscattering}
\end{align}
We are ignoring as in the main text any terms of order $\alpha \langle L_0\rangle$ or less.
In the low scattering limit the probability of scattering is $P_S \approx  \alpha \langle L_0 \rangle$.
Then the probability of a trapping event occurring is $P_S \bar{P}_{T,1}$.
The probability of no trapping is then $P_0=1-\alpha \langle L_0 \rangle \bar{P}_{T,1}\approx1$.
The probability of exactly one trapping event is $P_1=\alpha  \langle L_0 \rangle  \bar{P}_{T,1} (1-\bar{P}_{T,2})$.
Following this logic for $n$ trappings gives
\begin{align}
	P_n=\alpha \langle L_0 \rangle (1-\bar{P}_{T,n+1})\prod_{k=1}^{n} \bar{P}_{T,k} .
\end{align}
Plugging all this into Eq.~\ref{llowscattering}, using $\alpha l_\alpha=1$, and assuming that the probability of trapping converges to $\bar{P}_{T,\infty}$ as the number of scatterings approaches $\infty$:
\begin{align}
	\langle L\rangle &=  \langle L_0 \rangle + \langle L_0 \rangle (1-\bar{P}_{T,\infty}) \sum_{n=1}^\infty n\prod_{k=1}^n\bar{P}_{T,k}.
\end{align}

As expected, this expression reduces to Eq.~\Eqtrapped~in the main text when $\bar{P}_{T,n}=P_T$.

\section{Explicit integral calculation of $\langle L^0\rangle$ for a cube} \label{AppCube}

\begin{figure}
	\includegraphics[width=8.6cm]{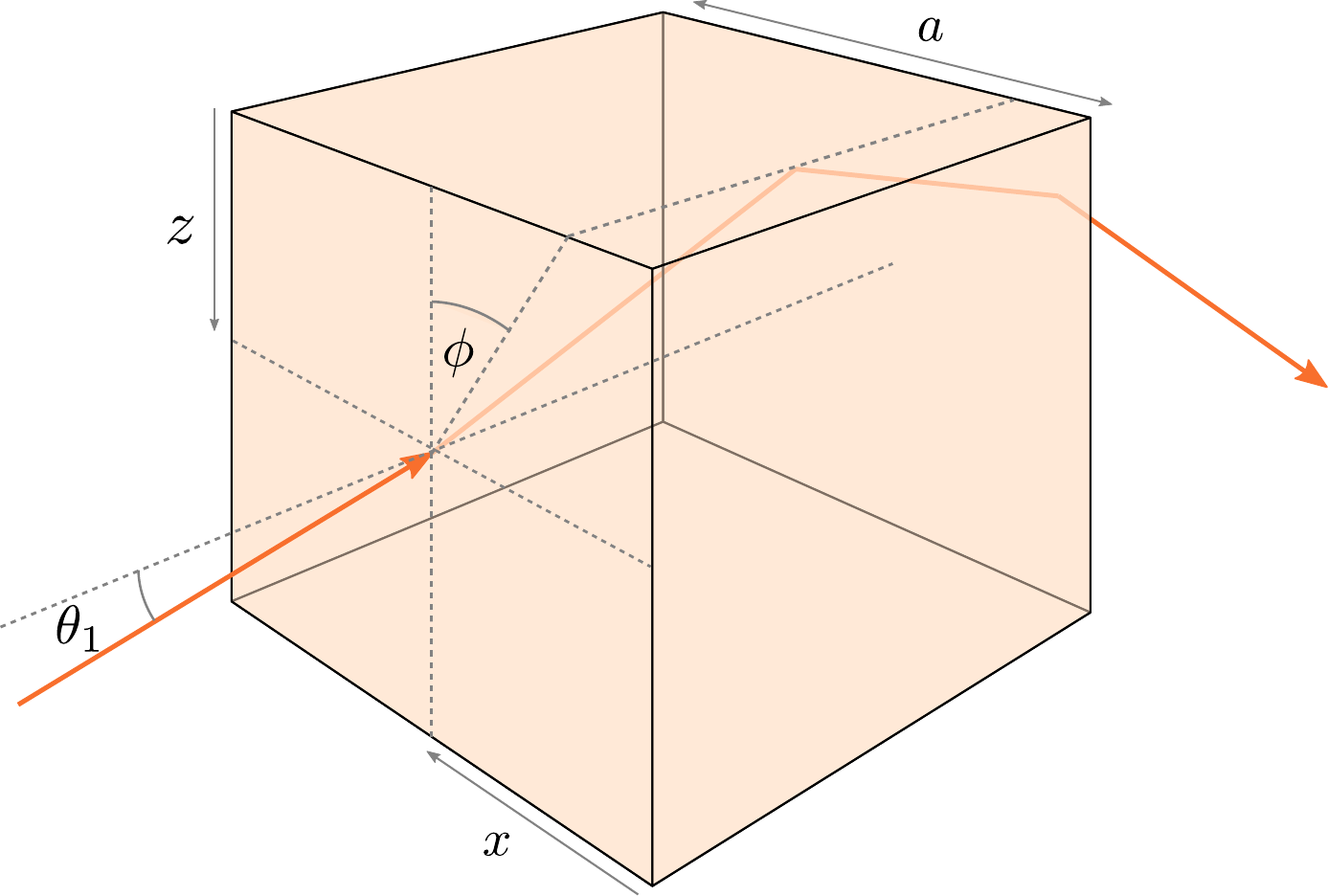}
	\caption{A light ray refracting into the ``front" face of a cube, reflecting off the top face, and refracting out the opposite face.} \label{fig cube}
\end{figure}

We start from Eq.~5 in the main text.
We integrate over the two angles of entry and over a vertically oriented face of the cube, which covers $0\leq z\leq 1$ and $0\leq x \leq 1$, where $z=0$ at the top and $x=0$ on the right, as per Fig.~\ref{fig cube}. The integral may be split into four:
\begin{align}
	\langle L_\text{cube}^0 \rangle = \frac{4 s^2}{\pi} (I_t+I_{o1}+I_{o2}+I_r), \label{cube integrals}
\end{align}
where $I_t$ counts the rays that refract towards the top face then refract out:
\begin{align}
	I_t=&\int_{a_1}^{b_1}\int_0^{\xi_{\rm TRE}}\int_{0}^{1}\int_0^1 L_t(\theta_2,\phi,z) p(\theta_2) \d z \d x \d \phi \d\theta_2 ,
\end{align}
$I_{o1}$ counts the rays that refract towards the top face then reflect towards the opposite face (depicted in Fig.~\ref{fig cube}), or leave the opposite face directly:
\begin{align}
	I_{o1}=&\int_{a_2}^{b_2}\int_0^{\xi_{\rm TRE}}\int_{0}^{1}\int_0^1 L_o(\theta_2) p(\theta_2) \d z \d x \d \phi \d\theta_2 ,
\end{align}
$I_{o2}$ counts the rays that refract towards the top face, reflect off both the top and right faces, and leave out the opposite face:
\begin{align}
	I_{o2}=&\int_{a_3}^{b_3}\int_0^{\xi_{\rm TRE}}\int_{0}^{1}\int_0^1 L_o(\theta_2) p(\theta_2) \d z \d x \d \phi \d\theta_2 ,
\end{align}
and $I_r$ counts the rays that refract towards the top face, reflect off the top face, and leave out the right face:
\begin{align}
	I_r=&\int_{a_4}^{b_4}\int_0^{\xi_{\rm TRE}}\int_{0}^{1}\int_0^1 L_r(\theta_2,\phi,x) p(\theta_2) \d z \d x \d \phi \d\theta_2. \label{cube integrals explicit}
\end{align}
$p(\theta_2)=2\cos\theta_2\sin\theta_2$, and the distances to the top face $(L_t)$, opposite face $(L_o)$ and right face $(L_r)$ are 
\begin{align*}
	L_o=&\frac{a}{\cos\theta_2}\\
	L_t=&\frac{z}{\sin\theta_2\cos\phi}\\
	L_r=&\frac{x}{\sin\theta_2\sin\phi}.
\end{align*}
$\phi$ is measured clockwise from the vertical as shown in Fig.~\ref{fig cube}. By symmetry we can assume that $\phi\ge 0$ and that all rays head towards the top face ($\phi<\xi_{\rm TRE}$), where 
\begin{align}
	\xi_{\rm TRE}=\tan^{-1}\dfrac{x}{z}
\end{align}
is the top-right edge.
The integral bounds for $\theta_2$ are
\begin{align*}
	a_1=&\min(\theta_c,\max(\xi_{\rm TOE},\xi_{\rm TIRT})) \\
	b_1=&\theta_c \\
	a_2=&0 \\ 
	b_2=&\min(\theta_c,\min(\xi_{\rm ORE},\max(\xi_{\rm TOE},\xi_{\rm TIRT})))\\
	a_3=&\min(\theta_c,\xi_{\rm ORE}) \\
	b_3=&\min(\theta_c,\max(\xi_{\rm ORE},\xi_{\rm TIRR}))\\
	a_4=&\min(\theta_c,\max(\xi_{\rm ORE},\xi_{\rm TIRR}))\\
	b_4=&\min(\theta_c,\max(\xi_{\rm ORE},\xi_{\rm TIRT})),
\end{align*}
where $\xi_{\rm TOE}$ is the top-opposite edge:
\begin{align}
	\xi_{\rm TOE}=\tan^{-1}\frac{z}{\cos\phi},
\end{align}
$\xi_{\rm ORE}$ is the opposite-right edge:
\begin{align}
	\xi_{\rm ORE}=\tan^{-1}\frac{x}{\cos\phi},
\end{align}
$\xi_{\rm TIRT}$ is the condition for total internal reflection off the top face:
\begin{align}
	\xi_{\rm TIRT}=\sin^{-1}\frac{\cos\theta_c}{\cos\phi},
\end{align}
$\xi_{\rm TIRR}$ is the condition for total internal reflection off the right face (this can only occur if the ray first undergoes TIR off the top face):
\begin{align}	
	\xi_{\rm TIRR}=\min\bigg(\xi_{\rm TIRT},\sin^{-1}\frac{\cos\theta_c}{\sin\phi}\bigg).
\end{align}
In order to evaluate these integrals analytically, they can be broken down into regions of $x,z,\phi$ where the max and min functions are not necessary. This results in about 50 4-dimensional integrals, most of which can be evaluated as sums of trigonometric and logarithmic functions and elliptic integrals.
We do not provide details here, but a similar simpler derivation is given in the case of the square in Sec.~\ref{App2D}.
On adding all integrals together the result simplifies to Eq.~\EqLcubeIntermediate. This remarkable simplification highlights the importance of the alternative approach based on trapping probability $P_T$, developed in the next section.

\section{Calculation of $\langle L ^0 \rangle$ for a cube using Eq.~\Eqtrapped}\label{AppCubeSimple}

We here use instead Eq.~\Eqtrapped, which applies because the trapping probability $P_T$ is independent of the location of the scattering event.
We can formulate the condition for these trapped rays by requiring TIR off all six faces.
Without loss of generality, let our trapped ray bounce off the front face with $\theta>\theta_c$ and some $\phi$. Then the condition for TIR off the top face (and also the bottom face) is $\sin\theta<\frac{\cos\theta_c}{\cos\phi}$. And similarly the condition for TIR off the right and left faces is $\sin\theta<\frac{\cos\theta_c}{\sin\phi}$. Finding trapped trajectories is equivalent to finding $\theta\geq \theta_c$ and $\phi$ fulfilling these two constraints.
The boundary of solutions lies at $\theta=\theta_c$. The other two conditions are satisfied if $\tan\theta_c$ is less than the maximum value that $1/\cos\phi$ and $1/\sin\phi$ can have simultaneously. This occurs at $\phi=\pi/4$ where $1/\cos\phi=1/\sin\phi=\sqrt{2}$. So trapped trajectories exist for $\tan\theta_c<\sqrt{2}$, or equivalently  $s>\sqrt{3/2}$. Moreover, the probability of trapping a ray is given by (using the 8-fold symmetry about $\phi$):
\begin{align}
	P_T=&\frac{2}{\pi}\int_{\cos^{-1}\cot\theta_c}^{\pi/4} \d\phi \int_{\theta_c}^{\sin^{-1}\tfrac{\cos\theta_c}{\cos\phi}}\sin\theta \d\theta.
\end{align}
Then Eq.~\Eqtrapped gives after integration:
\begin{align}
	&\langle L_\text{cube}^0(\sqrt{3/2}<s<\sqrt{2}) \rangle =\\
	&{\frac{4as^2}{\pi}}\left( \sin^{-1}(s^2-1) -  \sqrt{1-\frac{1}{s^2}}~\sin^{-1}(2s^2-3)\right). \label{LcubeIntermediate} \nonumber
\end{align}

\section{Infinite circular cylinder}\label{AppCylinder}

\begin{figure}
	\center{\includegraphics[width=5cm]{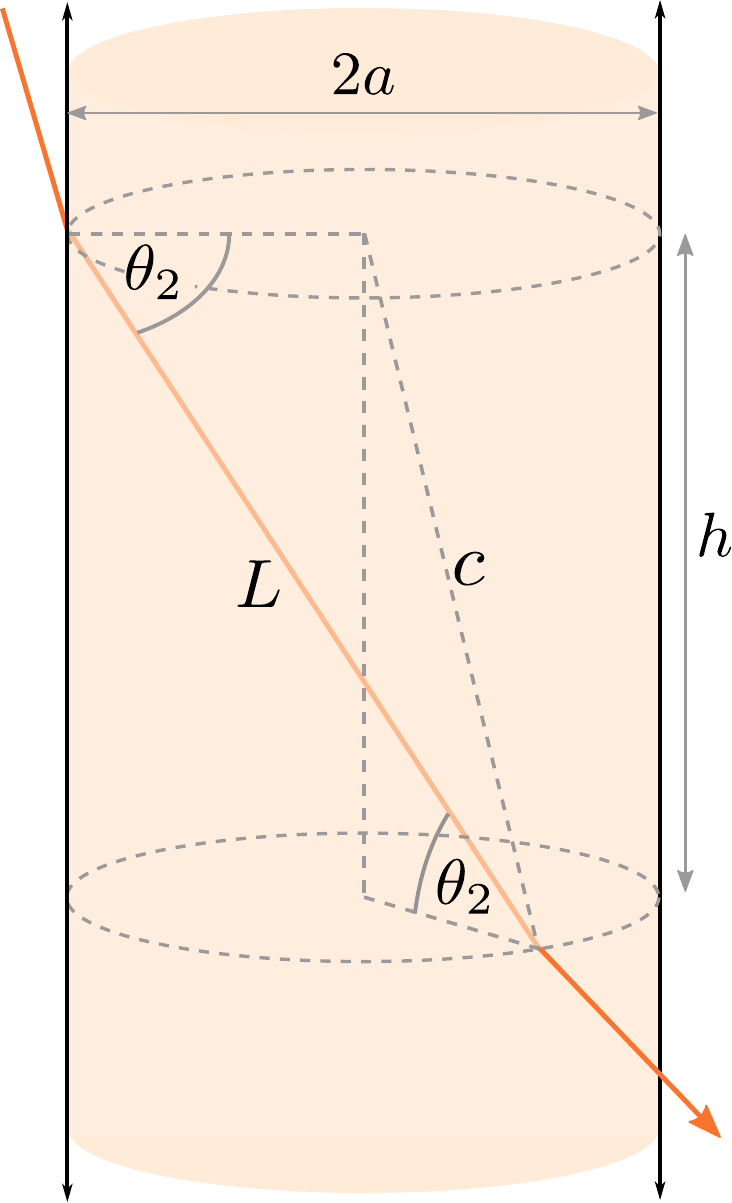}}
	\caption{A ray refracting through a cylinder. The angle $\phi$ is not shown as it comes out of the page; it is measured from the the vertical to the projection of the ray to the plane coming out of the page.} \label{Cylinder}
\end{figure}

For a cylinder of radius $a$, every incident point is equivalent, and the angle to the normal between internal bounces is invariant, a consequence of the reflectional and rotational symmetries of the cylinder. As a result, it is not possible to excite total internal reflections from the outside, and hence the mean path length and mean chord length are identical. 
The chord length $L$ can be derived from the cosine rule $c^2 = L^2+a^2-2 a L \cos\theta_2$,
where $c^2=h^2+a^2$, $h$ being the height along the cylinder that the ray travels; see Fig.~\ref{Cylinder}. $h$ is related to $L$ through $h=L\sin\theta_2\cos\phi$. Putting this together gives
\begin{align}
	L_{\rm cylinder} = \frac{2a\cos\theta_2}{1-\sin^2\theta_2\cos^2\phi},
\end{align}
and for the mean:
\begin{align}
	\langle L^0_\text{cylinder}\rangle =& \frac{s^2}{\pi}\int_0^{\theta_c}\int_0^{2\pi}  \frac{2a\cos^2\theta_2\sin\theta_2}{1-\sin^2\theta_2\cos^2\phi}  ~\d\phi \d\theta_2\nonumber \\
	=& 4as^2\int_0^{\theta_c} \sin\theta_2\cos\theta_2 ~ \d\theta_2\nonumber \\
	=&2a,
\end{align}
which, interestingly, is independent of $s$.
The cylinder qualitatively combines a slab along the axial direction and a circle along the radial one. The opposite effect of these shapes on $s$ seems to cancel perfectly. 
For comparison, the scattering mean path length is $\langle L_\text{cylinder}\rangle = 2a s^2$.
The discrepancy between the scattering and non-scattering cases may be attributed to trapped rays which run down the length of the cylinder.

\section{Infinite square rod}\label{AppRod}
%
%


For the {\it infinite square rod}, the derivations are similar to that of the cube and 2D square (see below) so are not repeated;
there are two regimes:  
\begin{align}
	&\langle L_\text{square rod}^0(s<\sqrt{2}) \rangle=\\
	&\frac{a s^2}{\pi}\left(\cos ^{-1}\left(1-s^2\right)-\sqrt{1-\frac{1}{s^2}}~ \cos ^{-1}\left(3-2s^2\right) \right),\nonumber
\end{align}
\begin{align}
	\langle L_\text{square rod}^0(s\geq\sqrt{2}) \rangle 
	=&as^2\bigg(1-\sqrt{1-\frac{1}{s^2}}\bigg).
\end{align}
These results are both less than the mean path length with scattering:
$\langle L_\text{square rod} \rangle = as^2$, due to trapped rays which run down the length of the rod.

\section{2D objects}\label{App2D}

We consider the problem of calculating the mean path length for a convex 2D object with refractive index and no scattering.
The diffuse external radiation creates a Lambertian incidence at all points on the surface, where the angles of rays incident are distributed in a 2D problem as $\tfrac{1}{2}\cos\theta_1$. 
As in the 3D case, the average path length is obtained from integrating over all possible incident points and angles:
\begin{align}
	\langle L_{2D}^0 \rangle &= \frac{1}{2P}\int_P \int_{-\pi/2}^{\pi/2}  L(\theta_1,r)  \cos\theta_1  \d\theta_1 \d r  
\end{align}
where $P$ is the object's perimeter, and $\int_P$ is the integral around the perimeter, so that $P=\int_P \d r$. 
Since $L$ is more naturally expressed in terms of $\theta_2$, it will be convenient to parametrize the integral as
\begin{align}
	\langle L_{2D}^0 \rangle&= \frac{s}{2P} \int_P \int_{-\theta_c}^{\theta_c}L(\theta_2,r) \cos\theta_2 \d\theta_2  \d r.\label{L2D}
\end{align}

\begin{figure}
	\includegraphics[width=8.6cm]{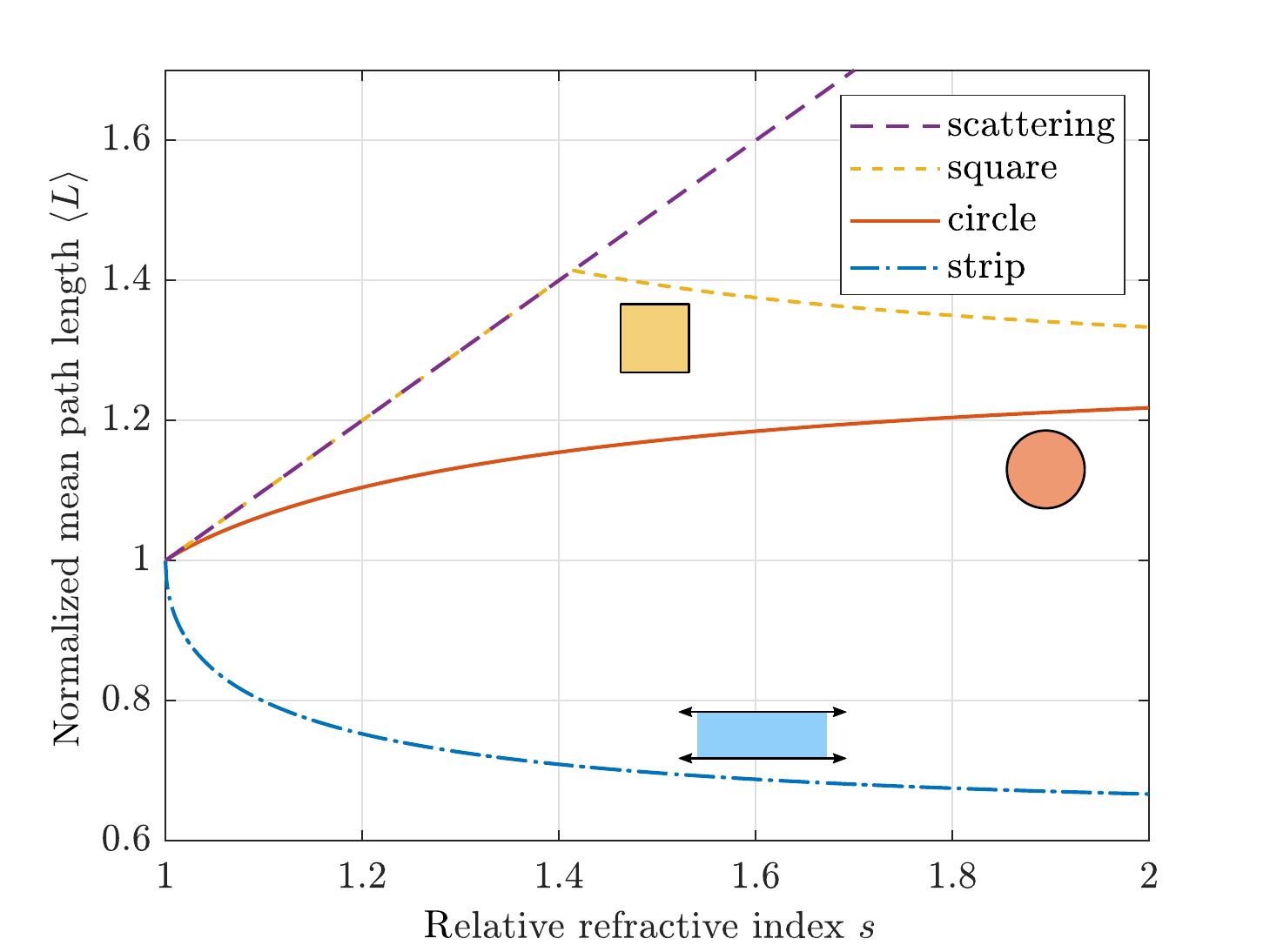}
	\caption{Comparison of the $s$ dependence of the zero-scattering mean path length $\langle L^0 \rangle$ for 2D objects for which analytical expressions were derived. All values are normalized to the mean chord length $\langle C \rangle=\pi A/P$.
		The scattering case $\langle L\rangle = s \langle C \rangle$ is shown as a dashed line.
	} \label{all2Dobjects}
\end{figure}

We derive analytic expressions in the following for simple 2D geometries, namely the infinite strip, cicle, and square.
The results are summarized and compared to the scattering case in Fig.~\ref{all2Dobjects}.

\subsection{Infinite strip}
We start with the simplest case of an infinite strip of width $a$. The derivation and results are very similar to the infinite slab presented in the main text.
The chord length is
\begin{align}
	L_{\rm strip} = \frac{a}{\cos\theta_2}.
\end{align}
Like the slab, there is no TIR and all surface points are identical. The integral \eqref{L2D} then reduces to 
\begin{align}
	\langle L_\text{strip}^0 \rangle &= \frac{as}{2}\int_{-\theta_c}^{\theta_c} \d\theta_2 \nonumber\\
	&=a s \theta_c . \label{lstrip}
\end{align}
This is less than the mean path length including scattering:
\begin{align}
	\langle L_\text{strip} \rangle =  a s\frac{\pi}{2}.
\end{align}
The result \eqref{lstrip} can also be obtained via Eq.~\Eqtrapped, since the probability of trapping is simply the fraction of angles greater than $\theta_c$: $P_T=1-2\theta_c/\pi$.

\subsection{Circle}
\begin{figure}
	\includegraphics[width=7cm]{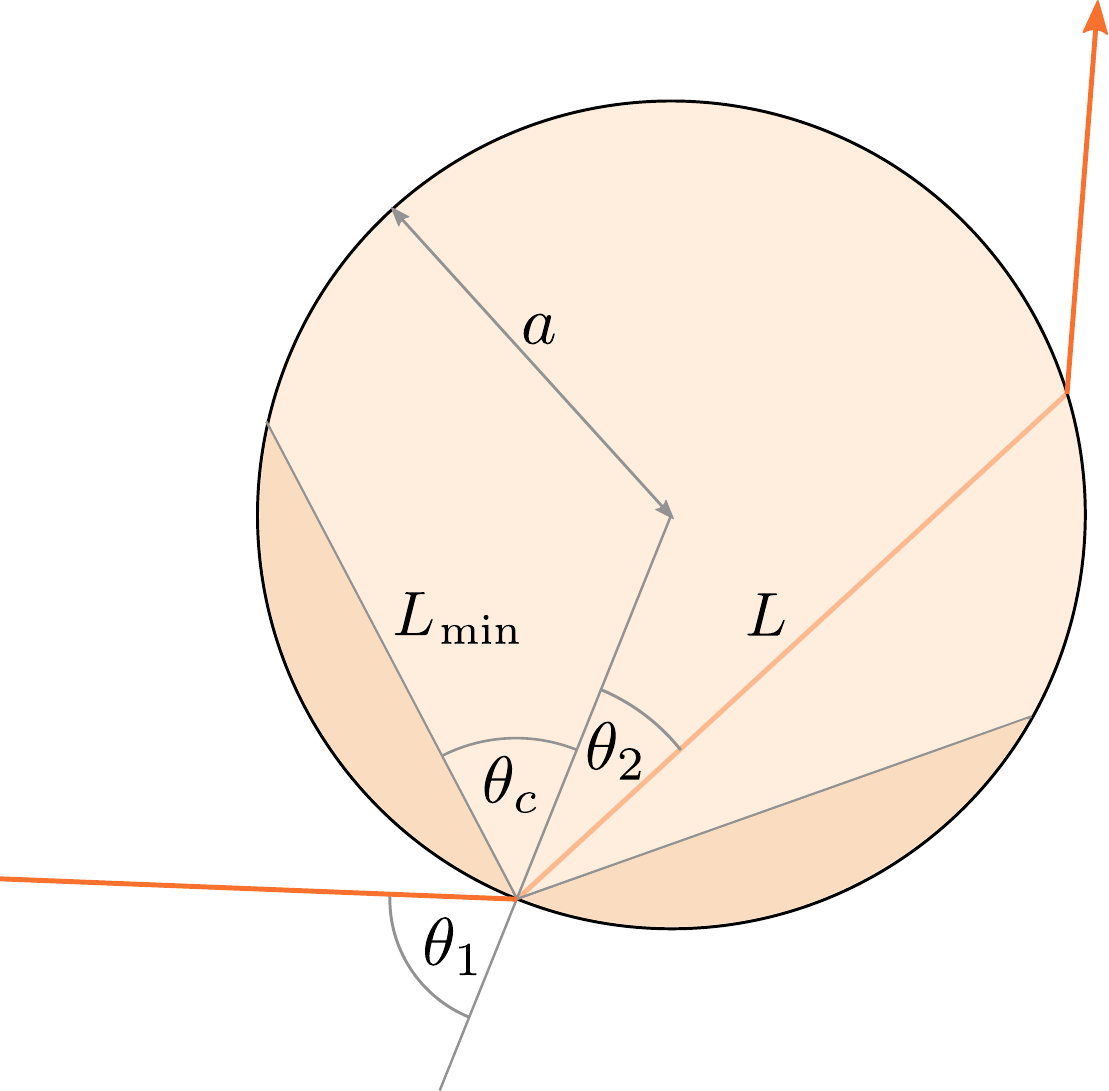}
	\caption{Light ray refracting through a circle. $L_{\rm min}$ is the minimum path length due to refraction.} \label{fig circle}
\end{figure}
For a circle of radius $a$, which is analogous to the 3D sphere in many respects, the chord length is again independent of the point of entry:
\begin{align}
	L_{\rm circle}=2a\cos\theta_2,
\end{align}
and the mean is then:
\begin{align}
	\langle L_\text{circle}^0 \rangle &= as \int_{-\theta_c}^{\theta_c} \cos^2\theta_2\d\theta_2 \nonumber\\
	&=a\left[ s \theta_c+ \cos\theta_c  \right], \label{lcircle}
\end{align}
which is less than the scattering mean path length:
\begin{align}
	\langle L_\text{circle} \rangle = \frac{a s}{2} .
\end{align}

\subsection{Square}

For a square of side length $a$ there are two cases, $s\le\sqrt{2}$ and  $s\geq\sqrt{2}$.
For $s\le \sqrt{2}$, then $\theta_c\geq \pi/4$ and there are no trapped rays; if a ray with angle $\theta_2$ undergoes TIR off one face - i.e. $\theta_2>\theta_c$ relative to that face's normal, then the ray then hits an adjacent side with an angle $\pi/2-\theta_2<\theta_c$.
The mean path length is therefore identical to that for the scattering case, i.e. 
\begin{align}
	\langle L_\text{square}^0(s\le \sqrt{2}) \rangle = \langle L_\text{square} \rangle = \frac{\pi}{4} a s. \label{Lsquare1}
\end{align}

For $s\geq\sqrt{2}$, since as explained in the main text reflections for $\theta<\theta_c$ may be ignored, the problem simplifies in that all rays that enter the square leave the opposite face, either due to hitting the opposite face directly, or via totally internally reflecting off an adjacent face, which is guaranteed if $\theta_c\le \pi/4$. Then $L=a/\cos\theta_2$ regardless of the incident angle or point of entry, and the integral \eqref{L2D} simplifies to
\begin{align}
	\langle L_\text{square}^0(s\geq\sqrt{2}) \rangle = a s \theta_c. \label{Lsquare2}
\end{align}
this expression coincides with \eqref{Lsquare1} when $s=\sqrt{2}$.\\

\begin{figure}[t]
	\center{\includegraphics[width=7cm]{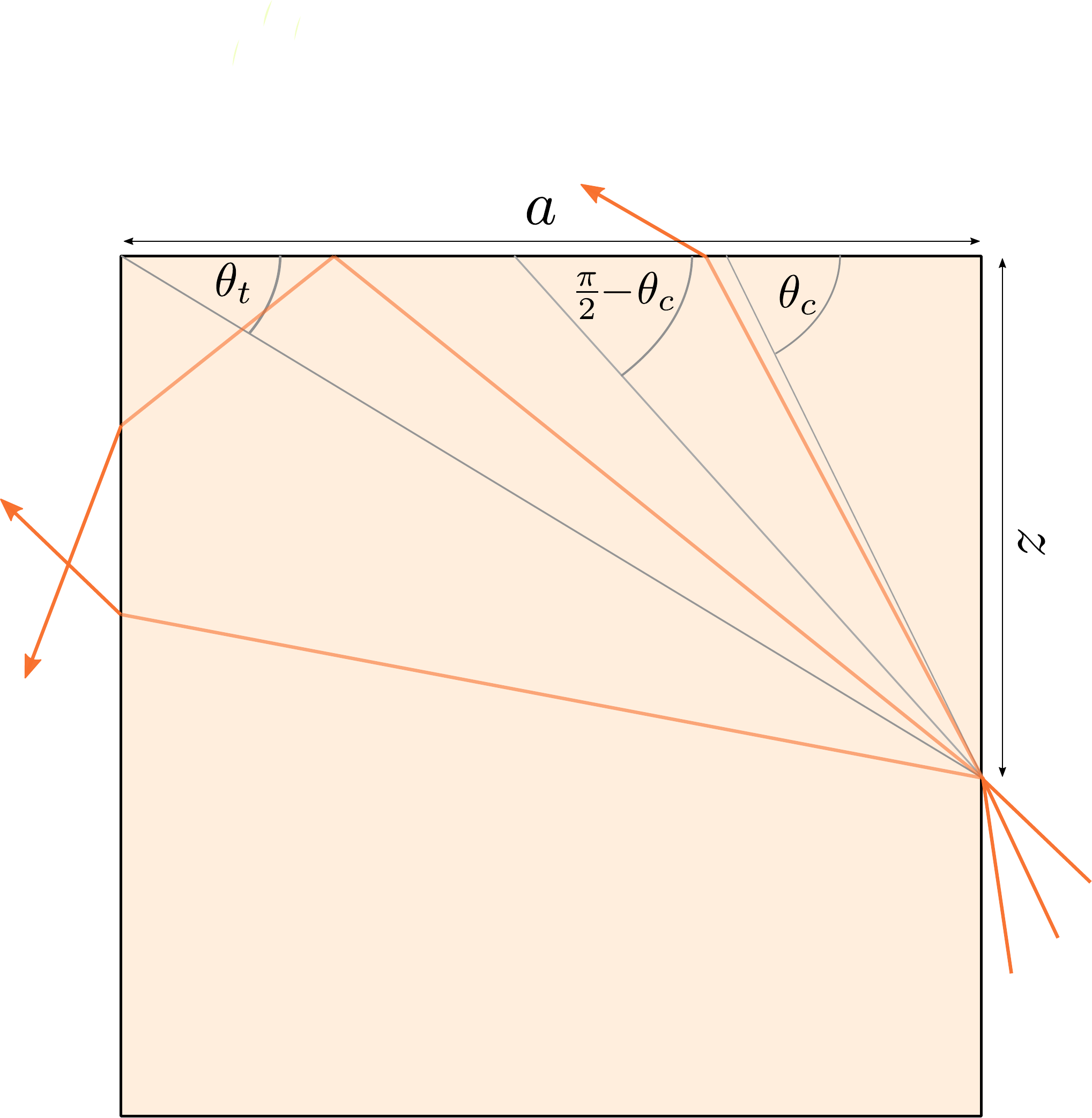}}
	\caption{Three different cases of light rays (in red) crossing a square, broken down into different angular regions. This square has $s\le \sqrt{2}$, ($\theta_c\geq \frac{\pi}{4}$).}  \label{fig square2}
\end{figure}

Alternatively, these results can be found, albeit with more difficulty, from the integral \eqref{L2D} for $s\le \sqrt{2}$ as follows. 
Consider a test ray entering from the right along $0\leq z\leq a$. By symmetry we may let $0\leq \theta_2\leq \theta_c$. 
As shown in Fig.~\ref{fig square2}, the ray can either hit the opposite face directly, with a length $L=a/\cos\theta_2$, hit the top face and TIR and leave out the opposite face (giving a total path of $L={a}/{\cos\theta_2}$ again), or hit the top face and leave with $L={z}/{\sin\theta_2}$.
This depends on the angle from the point of entry to the opposite top corner, $\theta_t=\tan^{-1}({z}/{a})$. Note that $\theta_t$ is always less than $\theta_c$, but $\theta_t$ may be greater or less than ${\pi}/{2}-\theta_c$. Explicitly:
\begin{align}
	L_{\rm square}=\left\{\begin{aligned}
		&\frac{a}{\cos\theta_2} \quad & 0\leq \theta_2\leq \theta_t,  \\
		&\frac{a}{\cos\theta_2}  & \theta_t\leq\theta_2\leq \max(\tfrac{\pi}{2}-\theta_c,\theta_t),\\
		&\frac{z}{\sin\theta_2} & \max(\tfrac{\pi}{2}-\theta_c,\theta_t)\leq\theta_2\leq\theta_c.
	\end{aligned}\right.
\end{align}
Inserting this into the integral \eqref{L2D} gives
\begin{align}
	\langle L_\text{square}^0(s\le\sqrt{2}) \rangle =&\frac{s}{a}\int_{0}^a \bigg[\int_0^{\max(\tfrac{\pi}{2}-\theta_c,\theta_t)} a\d \theta_2 \nonumber\\
	&+\int_{\max(\tfrac{\pi}{2}-\theta_c,\theta_t)}^{\theta_c} \frac{z}{\sin\theta_2}\cos\theta_2\d \theta_2  \bigg]\d z.
\end{align}
This can be evaluated analytically by splitting the integrals at the point $\theta_t=\pi/2-\theta_c$, which is equivalent to $z=a\cot\theta_c$. If $z<a\cot\theta_c$, then $\theta_t<\pi/2-\theta_c$, and vice versa. Then 
\begin{align}
	\langle L_\text{square}^0(s\le\sqrt{2}) \rangle =&\frac{s}{a}\bigg\{\int_{0}^{a\cot\theta_c} \bigg[\int_0^{\tfrac{\pi}{2}-\theta_c} a\d \theta_2 \nonumber\\
	&+\int_{\tfrac{\pi}{2}-\theta_c}^{\theta_c} \frac{z}{\sin\theta_2}\cos\theta_2\d \theta_2  \bigg]\d z \nonumber\\
	&+\int_{a\cot\theta_c}^a \bigg[\int_0^{\theta_t} a\d \theta_2 \nonumber\\
	&+\int_{\theta_t}^{\theta_c} \frac{z}{\sin\theta_2}\cos\theta_2\d \theta_2  \bigg]\d z\bigg\}.
\end{align}
The integrals individually evaluate to logarithmic and trigonometric functions, which together cancel to give $\langle L_\text{square}^0(s\le\sqrt{2}) \rangle=\pi as/4$.

\subsection{Rectangle}

For a rectangle of sides $a,b$, we can use the same arguments as for the square. For $s\le\sqrt{2}$ there are no trapped rays:
\begin{align}
	\langle L^0_\text{rect}(s<\sqrt{2})\rangle = \langle L_\text{rect}\rangle =\frac{\pi a b s }{2(a+b)}.
\end{align}
And for $s\geq\sqrt{2}$, again all rays exit the opposite face, so the integral \eqref{L2D} reduces to
\begin{align}
	\langle L^0_\text{rect}(s\geq\sqrt{2})\rangle = \frac{2a b s\theta_c}{a+b}.
\end{align}

\section{Absorption in a low scattering sample}
One important application of the study of path length is to study the optical absorption of a refractive object.
If we define the absorption $A$ as the fraction of rays that are absorbed compared to the total number of incident rays,
it can be expressed in terms of the path length distribution as \cite{2003BlancoEPL,2020TommasiPRA}:
\begin{align}
	A=1-\int_{0}^{\infty} e^{-\alpha_a L}p(L)\d L. \label{abs}
\end{align}
where $p(L)$ is the probability distribution of path length in the absence of absorption and $\alpha_a$ is the absorption coefficient.
In the low absorption limit, $\alpha_a\rightarrow0$, the exponential may be expanded in a Taylor series, giving approximately
\begin{align}
	A=\alpha_a \langle L\rangle + \cO(\alpha_a^2 \langle L\rangle ^2).\label{Aapprox}
\end{align}
This common approximation highlights the importance of the mean path length, but would also suggest that the measured absorption
could be discontinuous at zero scattering when $\langle L^0\rangle\neq \langle L\rangle$.
In practical situations ideal zero-scattering is not possible, so we here analyze the absorption in the limit of low scattering.
An implicit assumption in deriving Eq.~\eqref{Aapprox} is that there are no significant contributions in $p(L)$ from $L$ comparable or larger than $1/\alpha_a$,
since $\alpha_a L\ll 1$ is not valid for those $L$. 
This is no longer true in cases where there exists trapped, or even just long-lived, trajectories.


For objects with refractive index and a very low scattering coefficient $\alpha_s$, it is appropriate to split the path length distribution $p(L)$ into:
\begin{itemize}
	\item
	Rays that are directly reflected off the surface with probability $\bar{R}_{12}=1-\bar{T}_{12}$ (with zero path length), where $\bar{R}_{12}$ and $\bar{T}_{12}$ are the average (over a Lambertian distribution) reflection and transmission coefficients.
	\item
	Rays that enter, propagate inside, but do not scatter before exiting, with probability $P_0'$ and (renormalized) path length distribution $p_0(L)$.
	\item
	Rays that enter, scatter and exit without undergoing total internal reflection, with probability $P_S'$ and path length distribution $p_S(L)$.
	\item
	Rays that get trapped in long-lived trajectories, with probability $P_T'$ and path length distribution $p_T(L)$.
\end{itemize}
We can write explicitly:
\begin{align}
	p(L)= &\bar{R}_{12}\delta(L)+ P_0' p_0(L) +P_S' p_S(L) + P_T' p_T(L). \label{p low as}
\end{align}
The probabilities in the low scattering limit are, for a general object: 
\begin{align}
	P_0'=&\bar{T}_{12}-\alpha_s \langle L_0\rangle \nonumber\\
	P_S'=&(1-P_T)\alpha_s \langle L_0\rangle \nonumber\\
	P_T'=&\alpha_s \langle L_0\rangle P_T,
\end{align}
where $\langle L_0\rangle$ is the mean path length for zero-scattering and $P_T$ is the average probability of a scattered ray entering a trapped trajectory. Note that:
\begin{align}
	\bar{R}_{12}+P_0'+P_S'+P_T'=1.
\end{align}

To derive low order approximate expressions, the absorption may be split into contributions from different types of ray trajectories, following Eq.~\ref{p low as}:
\begin{align}
	A=P_0'A_0 + P_S' A_S + P_T' A_T,
\end{align}
where
\begin{align}
	A_0=1-\int_0^\infty p_0(L)e^{-\alpha_a L} \d L \nonumber\\
	A_S=1-\int_0^\infty p_S(L)e^{-\alpha_a L} \d L \nonumber\\
	A_T=1-\int_0^\infty p_T(L)e^{-\alpha_a L} \d L.
\end{align}
In the low absorbance limit, $A_0$ and $A_S$ are simplified by the fact that the path length distributions $p_0(L)$ and $p_S(L)$ are confined mostly to small $L$ relative to the absorption mean free path $1/\alpha_a$, so 
\begin{align}
	A_0\approx& \alpha_a \frac{\langle L_0\rangle}{\bar{T}_{12}} , \nonumber\\
	A_S\approx& \alpha_a \frac{\langle L_S\rangle}{\bar{T}_{12}} +\cO\Big(\frac{\alpha_s}{\alpha_a}\Big)
\end{align}
where $\langle L_0\rangle/\bar{T}_{12}$ is the mean path length of rays that enter, and $L_S/\bar{T}_{12}$ is the mean path length of rays that enter and scatter but do not get trapped into long-lived trajectories.

In order to simplify further, we now consider the limit of low absorption and \textit{lower} scattering, i.e. the scattering coefficient $\alpha_s$ is much less than the absorption coefficient, and both are small, $\alpha_s\ll\alpha_a\ll1/\langle L_0\rangle$.
$A_T$ can then be simplified to first order by recognizing that all trapped rays get absorbed since the absorption coefficient is much higher than the scattering coefficient, therefore:
\begin{align}
	A_T\approx1-\cO(\alpha_s/\alpha_a). 
\end{align}
Then altogether the absorption is 
\begin{align}
	A\approx& \bigg(1-\alpha_s\frac{\langle L_0\rangle }{\bar{T}_{12}}\bigg)  \alpha_a\langle L_0\rangle  \nonumber\\
	&+(1-P_T)\alpha_s\langle L_0\rangle \alpha_a\frac{\langle L_S\rangle}{\bar{T}_{12}}  \nonumber\\
	&+\alpha_s\langle L_0\rangle P_T.
\end{align}
In fact the terms containing the product $\alpha_a\alpha_s$ are second order, and neglecting them leaves
\begin{align}
	A&\approx \alpha_a \langle L_0\rangle +\alpha_s\langle L_0\rangle P_T \label{A1} \qquad\left[\alpha_s\ll\alpha_a\ll1/\langle L_0\rangle\right]. \nonumber\\
\end{align}
If $\alpha_s\ll \alpha_a$, it then reduces to $A\approx \alpha_a \langle L_0\rangle $.
This demonstrates that for realistic objects with residual scattering, the zero-scattering mean path length is the relevant quantity in terms of optical absorption, providing that absorption dominates over scattering. 
Although outside the scope of this work, the effect of other imperfections, such as the wave nature of light or surface scattering from imperfection, can be incorporated using similar arguments. In these cases we also deduce that a small non-zero absorption coefficient will dramatically limit the contribution of trapped and long-lived trajectories, resulting again in the zero-scattering mean path length being the relevant quantity.


\begin{thebibliography}{33}%
\makeatletter
\providecommand \@ifxundefined [1]{%
 \@ifx{#1\undefined}
}%
\providecommand \@ifnum [1]{%
 \ifnum #1\expandafter \@firstoftwo
 \else \expandafter \@secondoftwo
 \fi
}%
\providecommand \@ifx [1]{%
 \ifx #1\expandafter \@firstoftwo
 \else \expandafter \@secondoftwo
 \fi
}%
\providecommand \natexlab [1]{#1}%
\providecommand \enquote  [1]{``#1''}%
\providecommand \bibnamefont  [1]{#1}%
\providecommand \bibfnamefont [1]{#1}%
\providecommand \citenamefont [1]{#1}%
\providecommand \href@noop [0]{\@secondoftwo}%
\providecommand \href [0]{\begingroup \@sanitize@url \@href}%
\providecommand \@href[1]{\@@startlink{#1}\@@href}%
\providecommand \@@href[1]{\endgroup#1\@@endlink}%
\providecommand \@sanitize@url [0]{\catcode `\\12\catcode `\$12\catcode
  `\&12\catcode `\#12\catcode `\^12\catcode `\_12\catcode `\%12\relax}%
\providecommand \@@startlink[1]{}%
\providecommand \@@endlink[0]{}%
\providecommand \url  [0]{\begingroup\@sanitize@url \@url }%
\providecommand \@url [1]{\endgroup\@href {#1}{\urlprefix }}%
\providecommand \urlprefix  [0]{URL }%
\providecommand \Eprint [0]{\href }%
\providecommand \doibase [0]{https://doi.org/}%
\providecommand \selectlanguage [0]{\@gobble}%
\providecommand \bibinfo  [0]{\@secondoftwo}%
\providecommand \bibfield  [0]{\@secondoftwo}%
\providecommand \translation [1]{[#1]}%
\providecommand \BibitemOpen [0]{}%
\providecommand \bibitemStop [0]{}%
\providecommand \bibitemNoStop [0]{.\EOS\space}%
\providecommand \EOS [0]{\spacefactor3000\relax}%
\providecommand \BibitemShut  [1]{\csname bibitem#1\endcsname}%
\let\auto@bib@innerbib\@empty
\bibitem [{\citenamefont {Czuber}(1884)}]{1884Czuber}%
  \BibitemOpen
  \bibfield  {author} {\bibinfo {author} {\bibfnamefont {A.}~\bibnamefont
  {Czuber}},\ }\bibfield  {title} {\bibinfo {title} {Zur theorie der
  geometrischen wahrscheinlihkeiten},\ }\href@noop {} {\bibfield  {journal}
  {\bibinfo  {journal} {Sitzungsber. Akad. Wiss. Wien}\ }\textbf {\bibinfo
  {volume} {90}},\ \bibinfo {pages} {719} (\bibinfo {year} {1884})}\BibitemShut
  {NoStop}%
\bibitem [{\citenamefont {Kellerer}(1971)}]{1971Kellerer}%
  \BibitemOpen
  \bibfield  {author} {\bibinfo {author} {\bibfnamefont {A.~M.}\ \bibnamefont
  {Kellerer}},\ }\bibfield  {title} {\bibinfo {title} {Considerations on the
  random traversal of convex bodies and solutions for general cylinders},\
  }\href@noop {} {\bibfield  {journal} {\bibinfo  {journal} {Radiation Res.}\
  }\textbf {\bibinfo {volume} {47}},\ \bibinfo {pages} {359} (\bibinfo {year}
  {1971})}\BibitemShut {NoStop}%
\bibitem [{\citenamefont {Coleman}(1969)}]{coleman1969random}%
  \BibitemOpen
  \bibfield  {author} {\bibinfo {author} {\bibfnamefont {R.}~\bibnamefont
  {Coleman}},\ }\bibfield  {title} {\bibinfo {title} {Random paths through
  convex bodies},\ }\href@noop {} {\bibfield  {journal} {\bibinfo  {journal}
  {J. Appl. Proba.}\ ,\ \bibinfo {pages} {430}} (\bibinfo {year}
  {1969})}\BibitemShut {NoStop}%
\bibitem [{\citenamefont {De~Kruijf}\ and\ \citenamefont
  {Kloosterman}(2003)}]{de2003average}%
  \BibitemOpen
  \bibfield  {author} {\bibinfo {author} {\bibfnamefont {W.~J.~M.}\
  \bibnamefont {De~Kruijf}}\ and\ \bibinfo {author} {\bibfnamefont {J.~L.}\
  \bibnamefont {Kloosterman}},\ }\bibfield  {title} {\bibinfo {title} {On the
  average chord length in reactor physics},\ }\href@noop {} {\bibfield
  {journal} {\bibinfo  {journal} {Ann. Nucl. Energy}\ }\textbf {\bibinfo
  {volume} {30}},\ \bibinfo {pages} {549} (\bibinfo {year} {2003})}\BibitemShut
  {NoStop}%
\bibitem [{\citenamefont {Blanco}\ and\ \citenamefont
  {Fournier}(2003)}]{2003BlancoEPL}%
  \BibitemOpen
  \bibfield  {author} {\bibinfo {author} {\bibfnamefont {S.}~\bibnamefont
  {Blanco}}\ and\ \bibinfo {author} {\bibfnamefont {R.}~\bibnamefont
  {Fournier}},\ }\bibfield  {title} {\bibinfo {title} {An invariance property
  of diffusive random walks},\ }\href@noop {} {\bibfield  {journal} {\bibinfo
  {journal} {Europhys. Lett.}\ }\textbf {\bibinfo {volume} {61}},\ \bibinfo
  {pages} {168} (\bibinfo {year} {2003})}\BibitemShut {NoStop}%
\bibitem [{\citenamefont {Mupparapu}\ \emph {et~al.}(2015)\citenamefont
  {Mupparapu}, \citenamefont {Vynck}, \citenamefont {Svensson}, \citenamefont
  {Burresi},\ and\ \citenamefont {Wiersma}}]{2015MupparapuOE}%
  \BibitemOpen
  \bibfield  {author} {\bibinfo {author} {\bibfnamefont {R.}~\bibnamefont
  {Mupparapu}}, \bibinfo {author} {\bibfnamefont {K.}~\bibnamefont {Vynck}},
  \bibinfo {author} {\bibfnamefont {T.}~\bibnamefont {Svensson}}, \bibinfo
  {author} {\bibfnamefont {M.}~\bibnamefont {Burresi}},\ and\ \bibinfo {author}
  {\bibfnamefont {D.~S.}\ \bibnamefont {Wiersma}},\ }\bibfield  {title}
  {\bibinfo {title} {Path length enhancement in disordered media for increased
  absorption},\ }\href {https://doi.org/10.1364/OE.23.0A1472} {\bibfield
  {journal} {\bibinfo  {journal} {Opt. Express}\ }\textbf {\bibinfo {volume}
  {23}},\ \bibinfo {pages} {A1472} (\bibinfo {year} {2015})}\BibitemShut
  {NoStop}%
\bibitem [{\citenamefont {Tommasi}\ \emph
  {et~al.}(2020{\natexlab{a}})\citenamefont {Tommasi}, \citenamefont {Fini},
  \citenamefont {Martelli},\ and\ \citenamefont
  {Cavalieri}}]{tommasi2020invariance}%
  \BibitemOpen
  \bibfield  {author} {\bibinfo {author} {\bibfnamefont {F.}~\bibnamefont
  {Tommasi}}, \bibinfo {author} {\bibfnamefont {L.}~\bibnamefont {Fini}},
  \bibinfo {author} {\bibfnamefont {F.}~\bibnamefont {Martelli}},\ and\
  \bibinfo {author} {\bibfnamefont {S.}~\bibnamefont {Cavalieri}},\ }\bibfield
  {title} {\bibinfo {title} {Invariance property in scattering media and
  absorption},\ }\href@noop {} {\bibfield  {journal} {\bibinfo  {journal} {Opt.
  Comm.}\ }\textbf {\bibinfo {volume} {458}},\ \bibinfo {pages} {124786}
  (\bibinfo {year} {2020}{\natexlab{a}})}\BibitemShut {NoStop}%
\bibitem [{\citenamefont {Scheibelhofer}\ \emph {et~al.}(2018)\citenamefont
  {Scheibelhofer}, \citenamefont {Wahl}, \citenamefont {Larchev{\^e}que},
  \citenamefont {Chauchard},\ and\ \citenamefont
  {Khinast}}]{scheibelhofer2018spatially}%
  \BibitemOpen
  \bibfield  {author} {\bibinfo {author} {\bibfnamefont {O.}~\bibnamefont
  {Scheibelhofer}}, \bibinfo {author} {\bibfnamefont {P.~R.}\ \bibnamefont
  {Wahl}}, \bibinfo {author} {\bibfnamefont {B.}~\bibnamefont
  {Larchev{\^e}que}}, \bibinfo {author} {\bibfnamefont {F.}~\bibnamefont
  {Chauchard}},\ and\ \bibinfo {author} {\bibfnamefont {J.~G.}\ \bibnamefont
  {Khinast}},\ }\bibfield  {title} {\bibinfo {title} {Spatially resolved
  spectral powder analysis: experiments and modeling},\ }\href@noop {}
  {\bibfield  {journal} {\bibinfo  {journal} {Appl. Spectrosc.}\ }\textbf
  {\bibinfo {volume} {72}},\ \bibinfo {pages} {521} (\bibinfo {year}
  {2018})}\BibitemShut {NoStop}%
\bibitem [{\citenamefont {Sprafke}\ and\ \citenamefont
  {Wehrspohn}(2015)}]{2015Sprafke}%
  \BibitemOpen
  \bibfield  {author} {\bibinfo {author} {\bibfnamefont {A.~N.}\ \bibnamefont
  {Sprafke}}\ and\ \bibinfo {author} {\bibfnamefont {R.~B.}\ \bibnamefont
  {Wehrspohn}},\ }\bibinfo {title} {Current concepts for optical path
  enhancement in solar cells},\ in\ \href
  {https://doi.org/10.1002/9783527665662.ch1} {\emph {\bibinfo {booktitle}
  {Photon Management in Solar Cells}}}\ (\bibinfo  {publisher} {Wiley},\
  \bibinfo {year} {2015})\ Chap.~\bibinfo {chapter} {1}, pp.\ \bibinfo {pages}
  {1--20}\BibitemShut {NoStop}%
\bibitem [{\citenamefont {Sychugov}(2019)}]{sychugov2019analytical}%
  \BibitemOpen
  \bibfield  {author} {\bibinfo {author} {\bibfnamefont {I.}~\bibnamefont
  {Sychugov}},\ }\bibfield  {title} {\bibinfo {title} {Analytical description
  of a luminescent solar concentrator},\ }\href@noop {} {\bibfield  {journal}
  {\bibinfo  {journal} {Optica}\ }\textbf {\bibinfo {volume} {6}},\ \bibinfo
  {pages} {1046} (\bibinfo {year} {2019})}\BibitemShut {NoStop}%
\bibitem [{\citenamefont {Sychugov}(2020)}]{sychugov2020geometry}%
  \BibitemOpen
  \bibfield  {author} {\bibinfo {author} {\bibfnamefont {I.}~\bibnamefont
  {Sychugov}},\ }\bibfield  {title} {\bibinfo {title} {Geometry effects on
  luminescence solar concentrator efficiency: analytical treatment},\
  }\href@noop {} {\bibfield  {journal} {\bibinfo  {journal} {Appl. Opt.}\
  }\textbf {\bibinfo {volume} {59}},\ \bibinfo {pages} {5715} (\bibinfo {year}
  {2020})}\BibitemShut {NoStop}%
\bibitem [{\citenamefont {Wiersma}(2008)}]{2008WiersmaNP}%
  \BibitemOpen
  \bibfield  {author} {\bibinfo {author} {\bibfnamefont {D.~S.}\ \bibnamefont
  {Wiersma}},\ }\bibfield  {title} {\bibinfo {title} {The physics and
  applications of random lasers},\ }\href {https://doi.org/10.1038/nphys971}
  {\bibfield  {journal} {\bibinfo  {journal} {Nature Phys.}\ }\textbf {\bibinfo
  {volume} {4}},\ \bibinfo {pages} {359} (\bibinfo {year} {2008})}\BibitemShut
  {NoStop}%
\bibitem [{\citenamefont {Nelson}\ and\ \citenamefont
  {Pr\'{e}zelin}(1993)}]{1993NelsonAO}%
  \BibitemOpen
  \bibfield  {author} {\bibinfo {author} {\bibfnamefont {N.~B.}\ \bibnamefont
  {Nelson}}\ and\ \bibinfo {author} {\bibfnamefont {B.~B.}\ \bibnamefont
  {Pr\'{e}zelin}},\ }\bibfield  {title} {\bibinfo {title} {Calibration of an
  integrating sphere for determining the absorption coefficient of scattering
  suspensions},\ }\href {https://doi.org/10.1364/AO.32.006710} {\bibfield
  {journal} {\bibinfo  {journal} {Appl. Opt.}\ }\textbf {\bibinfo {volume}
  {32}},\ \bibinfo {pages} {6710} (\bibinfo {year} {1993})}\BibitemShut
  {NoStop}%
\bibitem [{\citenamefont {Villanueva}\ \emph {et~al.}(2016)\citenamefont
  {Villanueva}, \citenamefont {Veenstra},\ and\ \citenamefont
  {Steenbergen}}]{2016CalibrationAO}%
  \BibitemOpen
  \bibfield  {author} {\bibinfo {author} {\bibfnamefont {Y.}~\bibnamefont
  {Villanueva}}, \bibinfo {author} {\bibfnamefont {C.}~\bibnamefont
  {Veenstra}},\ and\ \bibinfo {author} {\bibfnamefont {W.}~\bibnamefont
  {Steenbergen}},\ }\bibfield  {title} {\bibinfo {title} {Measuring absorption
  coefficient of scattering liquids using a tube inside an integrating
  sphere},\ }\href {https://doi.org/10.1364/AO.55.003030} {\bibfield  {journal}
  {\bibinfo  {journal} {Appl. Opt.}\ }\textbf {\bibinfo {volume} {55}},\
  \bibinfo {pages} {3030} (\bibinfo {year} {2016})}\BibitemShut {NoStop}%
\bibitem [{\citenamefont {Ravey}\ and\ \citenamefont
  {Mazeron}(1982)}]{1982RaveyJO}%
  \BibitemOpen
  \bibfield  {author} {\bibinfo {author} {\bibfnamefont {J.-C.}\ \bibnamefont
  {Ravey}}\ and\ \bibinfo {author} {\bibfnamefont {P.}~\bibnamefont
  {Mazeron}},\ }\bibfield  {title} {\bibinfo {title} {Light scattering in the
  physical optics approximation; application to large spheroids},\ }\href@noop
  {} {\bibfield  {journal} {\bibinfo  {journal} {J . Opt.}\ }\textbf {\bibinfo
  {volume} {13}},\ \bibinfo {pages} {273} (\bibinfo {year} {1982})}\BibitemShut
  {NoStop}%
\bibitem [{\citenamefont {Chowdhury}\ \emph {et~al.}(1992)\citenamefont
  {Chowdhury}, \citenamefont {Barber},\ and\ \citenamefont
  {Hill}}]{chowdhury1992energy}%
  \BibitemOpen
  \bibfield  {author} {\bibinfo {author} {\bibfnamefont {D.~Q.}\ \bibnamefont
  {Chowdhury}}, \bibinfo {author} {\bibfnamefont {P.~W.}\ \bibnamefont
  {Barber}},\ and\ \bibinfo {author} {\bibfnamefont {S.~C.}\ \bibnamefont
  {Hill}},\ }\bibfield  {title} {\bibinfo {title} {Energy-density distribution
  inside large nonabsorbing spheres by using mie theory and geometrical
  optics},\ }\href@noop {} {\bibfield  {journal} {\bibinfo  {journal} {Appl.
  Opt.}\ }\textbf {\bibinfo {volume} {31}},\ \bibinfo {pages} {3518} (\bibinfo
  {year} {1992})}\BibitemShut {NoStop}%
\bibitem [{\citenamefont {Macke}(1993)}]{1993MackeAO}%
  \BibitemOpen
  \bibfield  {author} {\bibinfo {author} {\bibfnamefont {A.}~\bibnamefont
  {Macke}},\ }\bibfield  {title} {\bibinfo {title} {Scattering of light by
  polyhedral ice crystals},\ }\href@noop {} {\bibfield  {journal} {\bibinfo
  {journal} {Appl. Opt.}\ }\textbf {\bibinfo {volume} {32}},\ \bibinfo {pages}
  {2780} (\bibinfo {year} {1993})}\BibitemShut {NoStop}%
\bibitem [{\citenamefont {Bi}\ and\ \citenamefont
  {Yang}(2013)}]{bi2013physical}%
  \BibitemOpen
  \bibfield  {author} {\bibinfo {author} {\bibfnamefont {L.}~\bibnamefont
  {Bi}}\ and\ \bibinfo {author} {\bibfnamefont {P.}~\bibnamefont {Yang}},\
  }\bibfield  {title} {\bibinfo {title} {Physical-geometric optics hybrid
  methods for computing the scattering and absorption properties of ice
  crystals and dust aerosols},\ }in\ \href@noop {} {\emph {\bibinfo {booktitle}
  {Light Scattering Reviews 8}}}\ (\bibinfo  {publisher} {Springer},\ \bibinfo
  {year} {2013})\ pp.\ \bibinfo {pages} {69--114}\BibitemShut {NoStop}%
\bibitem [{\citenamefont {Kokhanovsky}\ and\ \citenamefont
  {Zege}(1995)}]{kokhanovsky1995local}%
  \BibitemOpen
  \bibfield  {author} {\bibinfo {author} {\bibfnamefont {A.~A.}\ \bibnamefont
  {Kokhanovsky}}\ and\ \bibinfo {author} {\bibfnamefont {E.~P.}\ \bibnamefont
  {Zege}},\ }\bibfield  {title} {\bibinfo {title} {Local optical parameters of
  spherical polydispersions: simple approximations},\ }\href@noop {} {\bibfield
   {journal} {\bibinfo  {journal} {Appl. Opt.}\ }\textbf {\bibinfo {volume}
  {34}},\ \bibinfo {pages} {5513} (\bibinfo {year} {1995})}\BibitemShut
  {NoStop}%
\bibitem [{\citenamefont {Sun}\ \emph {et~al.}(2017)\citenamefont {Sun},
  \citenamefont {Yang}, \citenamefont {Kattawar},\ and\ \citenamefont
  {Zhang}}]{sun2017physical}%
  \BibitemOpen
  \bibfield  {author} {\bibinfo {author} {\bibfnamefont {B.}~\bibnamefont
  {Sun}}, \bibinfo {author} {\bibfnamefont {P.}~\bibnamefont {Yang}}, \bibinfo
  {author} {\bibfnamefont {G.~W.}\ \bibnamefont {Kattawar}},\ and\ \bibinfo
  {author} {\bibfnamefont {X.}~\bibnamefont {Zhang}},\ }\bibfield  {title}
  {\bibinfo {title} {Physical-geometric optics method for large size faceted
  particles},\ }\href@noop {} {\bibfield  {journal} {\bibinfo  {journal} {Opt.
  Express}\ }\textbf {\bibinfo {volume} {25}},\ \bibinfo {pages} {24044}
  (\bibinfo {year} {2017})}\BibitemShut {NoStop}%
\bibitem [{\citenamefont {Ackerman}\ and\ \citenamefont
  {Stephens}(1987)}]{ackerman1987absorption}%
  \BibitemOpen
  \bibfield  {author} {\bibinfo {author} {\bibfnamefont {S.~A.}\ \bibnamefont
  {Ackerman}}\ and\ \bibinfo {author} {\bibfnamefont {G.~L.}\ \bibnamefont
  {Stephens}},\ }\bibfield  {title} {\bibinfo {title} {The absorption of solar
  radiation by cloud droplets: An application of anomalous diffraction
  theory},\ }\href@noop {} {\bibfield  {journal} {\bibinfo  {journal} {J.
  Atmosph. Sci.}\ }\textbf {\bibinfo {volume} {44}},\ \bibinfo {pages} {1574}
  (\bibinfo {year} {1987})}\BibitemShut {NoStop}%
\bibitem [{\citenamefont {Mitchell}(2000)}]{mitchell2000parameterization}%
  \BibitemOpen
  \bibfield  {author} {\bibinfo {author} {\bibfnamefont {D.~L.}\ \bibnamefont
  {Mitchell}},\ }\bibfield  {title} {\bibinfo {title} {Parameterization of the
  {M}ie extinction and absorption coefficients for water clouds},\ }\href@noop
  {} {\bibfield  {journal} {\bibinfo  {journal} {J. Atmosph. Sci.}\ }\textbf
  {\bibinfo {volume} {57}},\ \bibinfo {pages} {1311} (\bibinfo {year}
  {2000})}\BibitemShut {NoStop}%
\bibitem [{\citenamefont {Xu}\ \emph {et~al.}(2003)\citenamefont {Xu},
  \citenamefont {Lax},\ and\ \citenamefont {Alfano}}]{xu2003anomalous}%
  \BibitemOpen
  \bibfield  {author} {\bibinfo {author} {\bibfnamefont {M.}~\bibnamefont
  {Xu}}, \bibinfo {author} {\bibfnamefont {M.}~\bibnamefont {Lax}},\ and\
  \bibinfo {author} {\bibfnamefont {R.~R.}\ \bibnamefont {Alfano}},\ }\bibfield
   {title} {\bibinfo {title} {Anomalous diffraction of light with geometrical
  path statistics of rays and a {G}aussian ray approximation},\ }\href@noop {}
  {\bibfield  {journal} {\bibinfo  {journal} {Opt. Lett.}\ }\textbf {\bibinfo
  {volume} {28}},\ \bibinfo {pages} {179} (\bibinfo {year} {2003})}\BibitemShut
  {NoStop}%
\bibitem [{\citenamefont {van~de Hulst}(1981)}]{vandehulst1981}%
  \BibitemOpen
  \bibfield  {author} {\bibinfo {author} {\bibfnamefont {H.~C.}\ \bibnamefont
  {van~de Hulst}},\ }\href@noop {} {\emph {\bibinfo {title} {Light scattering
  by small particles}}}\ (\bibinfo  {publisher} {Dover},\ \bibinfo {address}
  {New York},\ \bibinfo {year} {1981})\BibitemShut {NoStop}%
\bibitem [{\citenamefont {Bohren}\ and\ \citenamefont
  {Huffman}(2008)}]{bohren2008absorption}%
  \BibitemOpen
  \bibfield  {author} {\bibinfo {author} {\bibfnamefont {C.~F.}\ \bibnamefont
  {Bohren}}\ and\ \bibinfo {author} {\bibfnamefont {D.~R.}\ \bibnamefont
  {Huffman}},\ }\href@noop {} {\emph {\bibinfo {title} {Absorption and
  scattering of light by small particles}}}\ (\bibinfo  {publisher} {John Wiley
  \& Sons},\ \bibinfo {year} {2008})\BibitemShut {NoStop}%
\bibitem [{\citenamefont {Kokhanovsky}\ and\ \citenamefont
  {Macke}(1997)}]{kokhanovsky1997integral}%
  \BibitemOpen
  \bibfield  {author} {\bibinfo {author} {\bibfnamefont {A.~A.}\ \bibnamefont
  {Kokhanovsky}}\ and\ \bibinfo {author} {\bibfnamefont {A.}~\bibnamefont
  {Macke}},\ }\bibfield  {title} {\bibinfo {title} {Integral light-scattering
  and absorption characteristics of large, nonspherical particles},\
  }\href@noop {} {\bibfield  {journal} {\bibinfo  {journal} {Appl. Opt.}\
  }\textbf {\bibinfo {volume} {36}},\ \bibinfo {pages} {8785} (\bibinfo {year}
  {1997})}\BibitemShut {NoStop}%
\bibitem [{\citenamefont {Gille}(1999)}]{gille1999small}%
  \BibitemOpen
  \bibfield  {author} {\bibinfo {author} {\bibfnamefont {W.}~\bibnamefont
  {Gille}},\ }\bibfield  {title} {\bibinfo {title} {The small-angle scattering
  correlation function of the cuboid},\ }\href@noop {} {\bibfield  {journal}
  {\bibinfo  {journal} {J. Appl. Cryst.}\ }\textbf {\bibinfo {volume} {32}},\
  \bibinfo {pages} {1100} (\bibinfo {year} {1999})}\BibitemShut {NoStop}%
\bibitem [{\citenamefont {Savo}\ \emph {et~al.}(2017)\citenamefont {Savo},
  \citenamefont {Pierrat}, \citenamefont {Najar}, \citenamefont {Carminati},
  \citenamefont {Rotter},\ and\ \citenamefont {Gigan}}]{2017SavoSCI}%
  \BibitemOpen
  \bibfield  {author} {\bibinfo {author} {\bibfnamefont {R.}~\bibnamefont
  {Savo}}, \bibinfo {author} {\bibfnamefont {R.}~\bibnamefont {Pierrat}},
  \bibinfo {author} {\bibfnamefont {U.}~\bibnamefont {Najar}}, \bibinfo
  {author} {\bibfnamefont {R.}~\bibnamefont {Carminati}}, \bibinfo {author}
  {\bibfnamefont {S.}~\bibnamefont {Rotter}},\ and\ \bibinfo {author}
  {\bibfnamefont {S.}~\bibnamefont {Gigan}},\ }\bibfield  {title} {\bibinfo
  {title} {Observation of mean path length invariance in light-scattering
  media},\ }\href {https://doi.org/10.1126/science.aan4054} {\bibfield
  {journal} {\bibinfo  {journal} {Science}\ }\textbf {\bibinfo {volume}
  {358}},\ \bibinfo {pages} {765} (\bibinfo {year} {2017})}\BibitemShut
  {NoStop}%
\bibitem [{\citenamefont {Tommasi}\ \emph
  {et~al.}(2020{\natexlab{b}})\citenamefont {Tommasi}, \citenamefont {Fini},
  \citenamefont {Martelli},\ and\ \citenamefont {Cavalieri}}]{2020TommasiPRA}%
  \BibitemOpen
  \bibfield  {author} {\bibinfo {author} {\bibfnamefont {F.}~\bibnamefont
  {Tommasi}}, \bibinfo {author} {\bibfnamefont {L.}~\bibnamefont {Fini}},
  \bibinfo {author} {\bibfnamefont {F.}~\bibnamefont {Martelli}},\ and\
  \bibinfo {author} {\bibfnamefont {S.}~\bibnamefont {Cavalieri}},\ }\bibfield
  {title} {\bibinfo {title} {Invariance property in inhomogeneous scattering
  media with refractive-index mismatch},\ }\href
  {https://doi.org/10.1103/PhysRevA.102.043501} {\bibfield  {journal} {\bibinfo
   {journal} {Phys. Rev. A}\ }\textbf {\bibinfo {volume} {102}},\ \bibinfo
  {pages} {043501} (\bibinfo {year} {2020}{\natexlab{b}})}\BibitemShut
  {NoStop}%
\bibitem [{\citenamefont {Yablonovitch}(1982)}]{yablonovitch1982statistical}%
  \BibitemOpen
  \bibfield  {author} {\bibinfo {author} {\bibfnamefont {E.}~\bibnamefont
  {Yablonovitch}},\ }\bibfield  {title} {\bibinfo {title} {Statistical ray
  optics},\ }\href@noop {} {\bibfield  {journal} {\bibinfo  {journal} {J. Opt.
  Soc. Am.}\ }\textbf {\bibinfo {volume} {72}},\ \bibinfo {pages} {899}
  (\bibinfo {year} {1982})}\BibitemShut {NoStop}%
\bibitem [{SI()}]{SI}%
  \BibitemOpen
  \href@noop {} {}\bibinfo {note} {See Supplemental Material at [URL will be
  inserted by publisher] for additional technical details.}\BibitemShut {Stop}%
\bibitem [{\citenamefont {Duntley}(1942)}]{duntley1942optical}%
  \BibitemOpen
  \bibfield  {author} {\bibinfo {author} {\bibfnamefont {S.~Q.}\ \bibnamefont
  {Duntley}},\ }\bibfield  {title} {\bibinfo {title} {The optical properties of
  diffusing materials},\ }\href@noop {} {\bibfield  {journal} {\bibinfo
  {journal} {J. Opt. Soc. Am.}\ }\textbf {\bibinfo {volume} {32}},\ \bibinfo
  {pages} {61} (\bibinfo {year} {1942})}\BibitemShut {NoStop}%
\bibitem [{\citenamefont {Berry}(1981)}]{1981BerryEJP}%
  \BibitemOpen
  \bibfield  {author} {\bibinfo {author} {\bibfnamefont {M.~V.}\ \bibnamefont
  {Berry}},\ }\bibfield  {title} {\bibinfo {title} {Regularity and chaos in
  classical mechanics, illustrated by three deformations of a circular
  'billiard'},\ }\href@noop {} {\bibfield  {journal} {\bibinfo  {journal} {Eur.
  J. Phys.}\ }\textbf {\bibinfo {volume} {2}},\ \bibinfo {pages} {91} (\bibinfo
  {year} {1981})}\BibitemShut {NoStop}%
\end{thebibliography}
\end{document}